# Long bone microanatomy in elephants: microstructural insights into gigantic beasts

Camille Bader[1]*, Remy Gilardet[2], Nicolas Rinder[1], Victoria Herridge[3], John. R. Hutchinson[4], Alexandra Houssaye[1]

[1]Département Adaptations du Vivant, UMR 7179, Mécanismes adaptatifs et Évolution (MECADEV) CNRS/Muséum national d'Histoire naturelle, Paris, France; camille.bader@edu.mnhn.fr, alexandra.houssaye@mnhn.fr

[2]Conservatoire d'espaces naturels Provence-Alpes-Côte d'Azur (CEN PACA), Le Cannet-des-Maures, France ; remy.gilardet@gmail.com

[3]School of Biosciences, University of Sheffield, Sheffield, United Kingdom; v.herridge@sheffield.ac.uk

[4]Structure and Motion Laboratory, Department of Comparative Biomedical Sciences, Royal Veterinary College, Hatfield, United Kingdom, JHutchinson@rvc.ac.uk

*Corresponding author

## Abstract

One of the greatest challenges of terrestrial locomotion is resisting gravity. The morphological adaptive features of the limb long bones of extant elephants, the heaviest living terrestrial animals, have previously been highlighted; however, their bone microanatomy remain largely unexplored. Here we investigate the microanatomy of the six limb long bones in *Elephas maximus* and *Loxodonta africana*, using comparisons of virtual slices as well as robustness analyses, to understand how they were adapted to heavy weight-bearing. We find that the long bones of elephant limbs display a relatively thick cortex and a medullary area almost entirely filled with trabecular bone. This trabecular bone is highly anisotropic with trabecular orientations reflecting the mechanical load distribution along the limb. The respective functional roles of the bones are reflected in their microanatomy through variations of cortical thickness distribution and main orientation of the trabeculae. We find microanatomical adaptations to heavy weight support that are common to other heavy mammals. Despite these shared characteristics, the long bones of elephants are closer to those of sauropods due to their shared columnar posture, which allows a relaxation of morphofunctional constraints, and thus relatively less robust bones with a thinner cortex than would be expected in such massive animals.

**Keywords:** proboscideans; bone microanatomy; functional morphology; scaling

# INTRODUCTION

Gravity constitutes the main constraint in terrestrial locomotion, exerting a downward force proportional to an animal's body mass (Biewener, 1989; Bertram & Biewener, 1990; Biewener & Patek, 2018). As body length doubles isometrically, body mass increases by a factor of 8 (Schmidt-Nielsen, 1984), meaning that larger animals contend with relatively greater gravitational constraints compared to smaller ones. Giant quadrupeds are particularly challenged: terrestrial gigantism evolved several times throughout the evolution of amniotes, leading to the emergence of giant species displaying multi-ton body masses (Depéret Charles, 1907; Raia *et al.*, 2012; Baker *et al.*, 2015; Bokma *et al.*, 2016; Hutchinson, 2021). Some of these taxa, characterized by musculoskeletal and physiological adaptations allowing them to accommodate their massive weight, are termed graviportal. A number of these adaptations are physiological, such as an increased blood pressure to ensure perfusion of the organs despite the size of the body (White & Seymour, 2014) or a lower metabolic rate allowing to withstand longer period without food intake (Christiansen, 2004). Other adaptations involve alterations to the morphology and proportions of bones, such as an increased robustness (i.e. thicker shaft for a given length) or a relative size reduction of the autopod (manus, pes) bones compared to the stylopod (humerus, femur) and zeugopod (radius, ulna, tibia, fibula) bones (Gregory, 1912; Osborn, 1929; Coombs, 1978). However, skeletal adaptations are not limited to the shape of the bones. Bone microanatomy (i.e., the organization of the bony tissues) provides crucial information on how animals were able to reach such massive forms: the internal structure of bones, including their density, cortical thickness, and arrangement of bone tissues, allow for inferences regarding the stresses that are placed on the bones. When subjected to mechanical loading, including heavy weight, bones undergo structural adaptations to better support the increased stress placed upon them (Ruff & Hayes, 1983; Turner, 1998; Ruimerman, 2005; Habib & Ruff, 2008; Nikander *et al.*, 2010; Doube *et al.*, 2011; Bishop *et al.*, 2018). Indeed, several studies have shown that graviportality is reflected in the inner organization of the bones: unlike most terrestrial vertebrates, graviportal species often lack a medullary cavity, resulting in a dense, compact bone structure adapted to weight-bearing and high load distribution (Wall, 1983; Houssaye *et al.*, 2016, 2018; Nganvongpanit *et al.*, 2017; Lefebvre *et al.*, 2023). Additionally, an increased bone density and a thicker cortical layer provide greater resistances to bending, torsional, and compression forces (Currey & Alexander, 1985; Oxnard, 1990, 1993; Houssaye *et al.*, 2016; Canoville *et al.*, 2022).

Limb bones play a pivotal role by providing structural support and facilitating movement, so that their anatomy is highly impacted by shifts in body mass during evolution (Hildebrand, 1982; Biewener, 1989;

Smuts & Bezuidenhout, 1993, 1994; Polly, 2007). In graviportal animals, limb long bones have been shown to display specific shapes (Smuts & Bezuidenhout, 1993, 1994; Mallet *et al.*, 2019; Lefebvre *et al.*, 2022; Bader *et al.*, 2023, 2024) and microanatomical variations (Wall, 1983; Houssaye *et al.*, 2016; Nganvongpanit *et al.*, 2017; Etienne, 2023; Lefebvre *et al.*, 2023) linked to body mass support. However, these adaptations can vary greatly, even between species of similar body mass (Mallet *et al.*, 2019; Etienne, 2023), indicating that graviportality can be expressed in different ways. Although elephants share some graviportal characteristics with other giant-bodied mammalian taxa such as rhinoceroses and hippopotamuses, they also display unique postural and locomotor adaptations that are reflected in their skeleton (Gambaryan, 1974; Christiansen, 2007; Kokshenev & Christiansen, 2010).

In proboscideans, gigantism was made possible partly due to the acquisition of columnar limbs. While unique among extant mammals, the distinctive columnar architecture of the limbs appeared first (in the Mesozoic era) in quadrupedal sauropods dinosaurs (Wilson & Sereno, 1998; Wilson & Carrano, 1999; Wilson, 2005), in which the specific orientation of the limbs (straight and almost orthogonal to the ground at rest) allowed an increased reliance on axial compression along the limbs, and thus an improved ability to support multiple tons of body mass without proportionally increasing the robustness of the bones (Hildebrand, 1982). This columnar architecture was also acquired in massive taxa from other mammal lineages (e.g. uintatheriids, diprotodons; Osborn, 1900; Gregory, 1912; Coombs, 1983). As expected in graviportal taxa, the long bones of elephant and rhino limb are filled with trabecular tissue, providing increased load transmission along the bones (Houssaye *et al.*, 2016; Nganvongpanit *et al.*, 2017; Etienne, 2023). Both rhinos and hippos exhibit a thick cortex, helping to support their heavy weight (Houssaye *et al.*, 2016; Etienne, 2023). While giant quadrupedal sauropods also display a medullary area filled with trabecular bone, they differ from rhinos and hippos in their cortical thickness, which is much thinner than would be expected for their huge weight (Lefebvre *et al.*, 2023). This difference in cortical thickness is related to their columnar stance, which improves load transmission along the bones and thus relaxes constraints favoring a thick cortex. Additionally, the columnar stance greatly reduces the stresses otherwise associated with a more flexed limb, such as potentially high stresses imposed on antigravity muscles in order to maintain an upright posture. The independent acquisition of columnar limbs in sauropods and proboscideans is reflected in the shape of their limb long bones. Whereas elephants display robust bones compared to non-graviportal taxa, they also display much thinner bones that would be expected for animals their size (Christiansen 2007, Bader et al. 2024), thus resembling the pattern of decreased robustness observed in some sauropods (Lefebvre *et al.*, 2022). As a result, we would expect to observe shape and microanatomical patterns and variations in elephants that are closer to those of columnar-limbed sauropods, instead of those observed in more closely related and giant-bodied but relatively non-columnar mammalian taxa.

Extant elephants are the last surviving representatives of the clade Proboscidea, which includes numerous extinct giant taxa (e.g., *Gomphotherium*, *Mammut*, *Deinotherium*) (Sukumar, 2003; Larramendi, 2015). The three extant elephant species are divided into the genus *Elephas*, represented by the Asian elephant (*Elephas maximus*), and the African genus *Loxodonta*, comprising the African savanna elephant (*Loxodonta africana*) and the African forest elephant (*Loxodonta cyclotis*) (Grubb et al., 2020; Roca et al., 2001, 2004). While the three species share an overall similar morphology, they show significant variation in body dimensions and mass, and live in different habitats (closed and humid vs. open and arid). While certain characteristics like the shape of the spine or the autopod can differentiate between Asian and African elephant species (Wittemyer, 2011), limb long bones show very few morphological features allowing for species distinction (West, 2006; Todd, 2010): the main observable differences between *E. maximus* and *L. africana* lie in the robustness of the humerus, ulna and tibia, as well as in the enlargement and orientation of certain muscle attachments such as the olecranon tuberosity of the ulna or the tibial cranial crest (Bader *et al.*, 2023). Bone external morphology and microanatomy are complementary (Currey, 2002): different distributions can be observed between the two, with microanatomical features sometimes showing differences where the external morphology does not. As a result, we could expect to observe corresponding interspecific variations in the microanatomy of the humerus, ulna and tibia. Conversely, even if long bone shape does not reflect the weight and habitat difference between these two species, we could expect to observe microanatomical changes in the other three bones (radius, femur and fibula), allowing specific adaptations to weight-bearing or locomotion on different substrates.

The microanatomy and external morphology of bones may vary significantly depending on the specific bone and its location within the body. While they share a function of weight support, the six long bones do not participate equally in the support and the movement of the body, so that they face different biomechanical constraints (Shindo & Mori, 1956a,b; Bertram & Biewener, 1992; Smuts & Bezuidenhout, 1993, 1994) that might be expressed in different ways. As in most quadrupedal mammals, in elephants the weight is unequally distributed between limbs, with a greater percentage being supported by the forelimbs (Lessertisseur & Saban, 1967; Hildebrand, 1982; Pandy *et al.*, 1988; Polly, 2007; Ren *et al.*, 2010; Etienne *et al.*, 2021) where we expect to observe stronger adaptation to weight support. In addition, due to their position closer to the thorax and their involvement in the shoulder and hip joints respectively, the stylopods are subjected to more varying muscular constraints than the zeugopod (Smuts & Bezuidenhout, 1993, 1994; Shil *et al.*, 2013), so that we could expect a different pattern of microanatomical organization between the limb segments. Conversely, the zeugopod is subjected to simpler but stronger constraints: it is oriented more orthogonally to the ground so that it is subjected to relatively more axial compression (Hildebrand, 1982; Currey, 2002), although the difference is reduced in columnar graviportal taxa.

Bone microanatomy is also modified during ontogeny, as it undergoes changes to accommodate bone growth, increasing mass and changing functional needs (Enlow, 1963; Ricqles, 1991; Currey, 2002; Curtin *et al.*, 2012). In Asian elephants, stylopod bones display an ontogenetic allometry of shape, i.e. their adult morphology is different to that of juveniles. This variation is expressed through an increased growth of the proximal epiphyses as compared to the distal ones. Conversely, the zeugopod bones appear to follow an isometric growth pattern (Bader *et al.*, 2023). Adult elephants weigh between 30 and 90 times their birth weight (Wittemyer, 2011), which imply large variations in biomechanical constraints. We thus expect to observe varying degrees of microanatomical variation during growth depending on the bones considered.

Few studies have investigated the microanatomy of elephantid bones: based on mid-diaphyseal cross-sections of the femur and humerus, Houssaye et al. (2016) highlighted similarities in the bone deposition between graviportal taxa and some aquatic taxa. Only Nganvongpanit et al. (2017) used longitudinal cross-sections of the six long bones, providing information on the general cortical distribution although without describing the distribution of the trabecular bone. The details of the inner organization of elephants' bones thus remain largely unexplored, especially considering the interspecific and intraspecific variations.

In this study, we aim to: 1) describe the general pattern of limb bone microanatomy in the six long bones in extant elephants and investigate the patterns of microanatomical variations between limb segments, i.e. forelimb/hindlimb and stylopod/zeugopod; 2) explore the intraspecific and relative interspecific variations and to place these descriptions into the more general context of bone microanatomy among graviportal animals; and 3) compare these findings to what is known of the corresponding external modifications of shape in order to better understand the coupled adaptation of external morphology and microanatomy in the long bones of elephant limbs.

# MATERIAL & METHODS

## Sample & X-ray microtomography

We selected a total of 100 bones from 26 proboscidean specimens from several European institutions, belonging to two species, *Elephas maximus* and *Loxodonta africana*. Our sample for microanatomical analyses was composed of 19 humeri, 13 radii, 16 ulnae, 23 femora, 16 tibiae and 13 fibulae (Table 1).

Species determination was provided by the institutions. Age determination was sometimes available; otherwise, the ontogenetic stage was assumed based on the level of fusion and development of the epiphyses (calf: absent or unfused epiphyses, juvenile: partially fused epiphyses, subadult/adult: almost fully fused epiphyses/fully fused epiphyses; Table 1) (Roth, 1984; Herridge, 2010). The sex, origin and captivity state were generally unknown and could thus not be accounted for in our analyses.

A subsample of the bones was scanned using high-resolution computed tomography. 16 of them were scanned at the AST-RX platform (UMS 2700, Muséum National d'Histoire Naturelle, Paris; GE phoenix|X-ray v|tome|xs 240) with reconstructions performed using X-Act (RX-Solutions). Voxel size varied from 65 µm to 148 µm depending on specimen size. Another subsample (84 bones) was scanned using medical computed tomography (Picker PQ5000; Philips Healthcare, Andover, MA, USA) at the Royal Veterinary College's Equine Diagnostic Unit. Voxel size varied from 690 µm to 5000 µm, providing low resolution scans that were used only for robustness calculations as well as for qualitative comparisons. In addition, 133 3D models of bones that were used in previous studies on external morphology (Bader et al. 2023, 2024) were added for quantitative analyses of bone robustness. Bone tissues were segmented for the complete bones, using Avizo 9 (VSG, Burlington, MA, USA) or VGStudio MAX (2016, v. 2.2, Volume Graphics Inc.) (Table 1), depending on the size of the data.

**Table 1**. Sample studied. AM, acquisition mode; medCT, medical CT-scan; µCT, micro CT-scan. OS, ontogenetic stage, A, adult, J, juvenile, C, calf. Institutional codes: MNHN, Muséum national d'Histoire naturelle, Paris (France); NHMUK, Natural History Museum London (UK); RVC, Royal Veterinary College (UK). Abbreviations: Fe, femur; Fi, fibula; H, humerus; na, not available; R, radius; T, tibia; U, ulna.

| **Taxon** | **Specimen number** | **Institution** | **H** | **R** | **U** | **Fe** | **T** | **Fi** | **AM** | **Range (µm)** | **OS** |
|---|---|---|---|---|---|---|---|---|---|---|---|
| *Elephas maximus* | ZM-AC-1883-1786 | MNHN | | | | X | | | µCT | 65-148 | A |
| *Elephas maximus* | ZM-AC-1936-280 | MNHN | | | | | X | X | µCT | 65-148 | A |
| *Elephas maximus* | ZM-AC-1983-082 | MNHN | | | | | X | X | µCT | 65-148 | A |
| *Loxodonta africana* | ZM-AC-1907-49 | MNHN | | | | X | X | X | µCT | 65-148 | A |
| *Loxodonta africana* | ZM-AC-1938-375 | MNHN | | X | | | | | µCT | 65-148 | A |
| *Loxodonta africana* | ZM-AC-1977-030G | MNHN | | | | X | | | µCT | 65-148 | A |
| *Loxodonta africana* | ZM-AC-1986-060 | MNHN | X | | | | | | µCT | 65-148 | A |
| *Loxodonta africana* | ZM-AC-1961-069 | MNHN | X | X | X | X | X | | µCT | 65-148 | C |
| *Elephas maximus* | NHMUK-1907.3.18.1 | NHMUK | X | X | X | X | X | X | medCT | 690-5000 | A |
| *Elephas maximus* | RVC-GNR | RVC | X | | | X | | | medCT | 690-5000 | A |
| *Elephas maximus* | RVC-GTA | RVC | X | | | X | | | medCT | 690-5000 | A |
| *Elephas maximus* | NHMUK-1851.11.10.16 | NHMUK | X | X | X | X | X | | medCT | 690-5000 | C |
| *Elephas maximus* | NHMUK-1915.5.1.1 | NHMUK | X | X | X | X | X | X | medCT | 690-5000 | C |
| *Elephas maximus* | NHMUK-H.4611 | NHMUK | | | | | X | X | medCT | 690-5000 | C |
| *Elephas maximus* | NHMUK-H.4644 | NHMUK | X | | | X | | | medCT | 690-5000 | C |
| *Elephas maximus* | NHMUK-H.4647 | NHMUK | | | | | X | X | medCT | 690-5000 | C |
| *Elephas maximus* | NHMUK-Z1425 | NHMUK | X | | | | | | medCT | 690-5000 | C |
| *Elephas maximus* | NHMUK-Z1578 | NHMUK | | | X | | | | medCT | 690-5000 | C |
| *Elephas maximus* | NHMUK-Z1580-b | NHMUK | | | | X | | | medCT | 690-5000 | C |
| *Elephas maximus* | NHMUK-Z1580-c | NHMUK | | | | X | | | medCT | 690-5000 | C |
| *Elephas maximus* | NHMUK-1984.516 | NHMUK | X | X | X | X | X | X | medCT | 690-5000 | J |
| *Loxodonta africana* | NHMUK-1929.1.1.36 | NHMUK | | | | X | | | medCT | 690-5000 | A |
| *Loxodonta africana* | NHMUK-1939.5.27.1 | NHMUK | X | X | X | X | X | X | medCT | 690-5000 | A |
| *Loxodonta africana* | RVC-RSA | RVC | X | | | X | | | medCT | 690-5000 | A |
| *Loxodonta africana* | NHMUK-1918.5.22.2 | NHMUK | X | | X | X | | | medCT | 690-5000 | C |
| *Loxodonta africana* | NHMUK-1961.8.9.81 | NHMUK | X | X | X | X | X | | medCT | 690-5000 | C |
| *Loxodonta africana* | NHMUK-1962.7.6.8 | NHMUK | X | X | X | X | X | X | medCT | 690-5000 | C |
| *Loxodonta africana* | NHMUK-1962.8.17.2 | NHMUK | X | X | X | X | X | X | medCT | 690-5000 | C |
| *Loxodonta africana* | NHMUK-1984.514 | NHMUK | X | X | X | X | X | | medCT | 690-5000 | C |
| *Loxodonta africana* | NHMUK-1961.8.9.82 | NHMUK | X | X | X | X | X | X | medCT | 690-5000 | J |
| *Loxodonta africana* | NHMUK-1962.7.6.9 | NHMUK | X | X | X | X | X | X | medCT | 690-5000 | C |

## Qualitative analyses

Virtual cross-sections of the bones were created using VGStudio Max. Each bone was oriented in anatomical position following Smuts & Bezuidenhout (1993, 1994). Three perpendicular section planes were defined: transverse, coronal and sagittal (Fig. 1). We defined the transverse section as being perpendicular to the longitudinal axis of the bone and cutting through its center of ossification. The coronal and sagittal sections were defined as cutting through the medullary area. Due to their morphological specificities, additional cross-sections were added for the ulna and the fibula: the olecranon tuberosity of the ulna is massive and is medial to the medullary area, so we added a second cross-section coronally and sagittally, placed more medially. In proboscideans, the fibula is typically long and thin, with a variably 'twisted' diaphysis, so that a single cross-section cannot always cut through the entire bone along the longitudinal axis. A second coronal and a second sagittal cross-section were added for the fibulae, defined as cutting through the medullary space in the proximal and distal parts of the diaphysis.

Additionally, in order to facilitate visualization of cortical thickness variation, we created 3D-mappings on the long bones of juvenile and adult *Loxodonta africana* specimens (Fig. S13, Material S1).

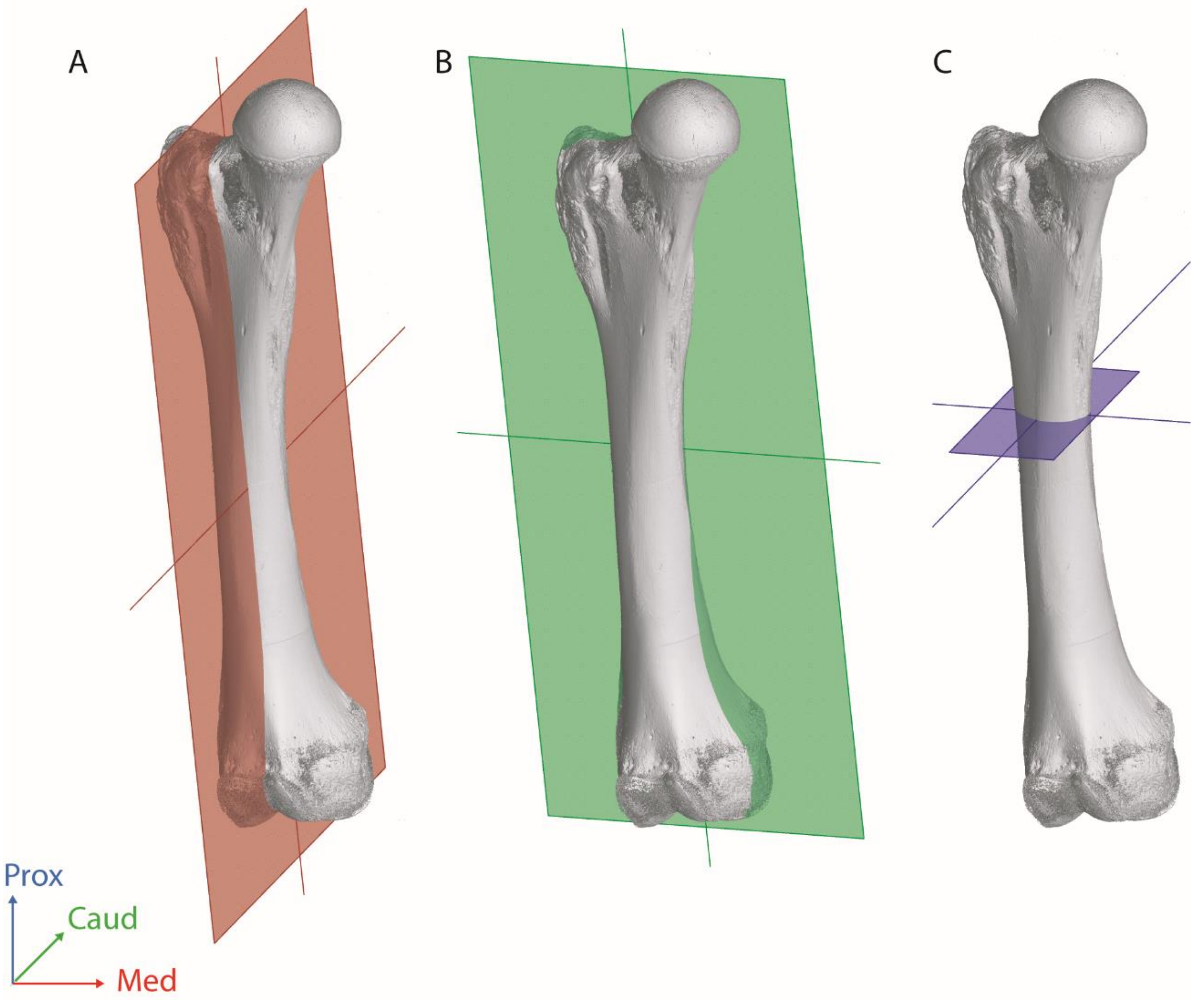

**Figure 1**. Representations of the virtual slices made in the femora. A: sagittal slice, B: coronal slice, C: transverse slice. Prox, proximal; Caud, caudal; Med, medial.

## Quantitative parameters

In order to assess the robustness of the long bones, we measured the maximal length and minimum diaphyseal circumference of each bone. Since our sample includes a large number of juvenile specimens, thus often lacking epiphyses, we chose to measure the maximal length of the diaphysis as a proxy of the entire bone maximal length (*MaxL*). *MaxL* was defined on each of the six bones as the distance between the proximal and the distal limits of the diaphysis as defined by specific homologous landmarks (Figs. S1-S6). The minimal diaphyseal circumference (*Ci*) was defined as the shortest circumference measurable on the diaphysis of the bone. *MaxL* and *Ci* were obtained digitally by using CloudCompare (version 2.12.0, http://www.cloudcompare.org). Robustness (*Rb*) was defined as the ratio of minimal diaphyseal circumference to maximal length of the bones (*Ci/MaxL*). We tested for differences in robustness between species and ontogenetic stages with ANOVAs associated with pairwise comparisons. We applied False Discovery Rate (FDR) correction (Benjamini & Hochberg 1995; Curran-Everett 2000) to control for false positives and ensure the robustness of the results.

# RESULTS

## Qualitative descriptions

In adults of both *Elephas maximus* and *Loxodonta africana*, the primary ossification center or growth center (GC), i.e. the first area of a bone that starts ossifying during growth (Carter & Beaupré, 2001), is located in the distal half of the diaphysis, as evident in sagittal and coronal views (Fig. 2A, B). The cortex, the thickest around the growth center, becomes thinner in both the proximal and distal directions so that the medullary area forms an asymmetrical "hourglass" shape; this is the case for all long bones except the fibula. Cortical thickness is overall similar in the two species; the relatively wider bones of *E. maximus* (compared to *L. africana*) are thus associated with a relatively larger medullary area. For all bones, in adults of both *E. maximus* and *L. africana* the delimitation between the cortex and the medullary area is clear throughout the diaphysis, whereas juveniles exhibit an imprecise distinction between the cortex and the medullary area due to the trabecular bone forming thick struts along the whole diaphysis.

### Humerus

In the humerus, the hourglass shape is much longer in the proximal half than in the distal one. However, in *E. maximus* (n=3), the medullary area is more cylindrical, the cortical bone forming a less marked hourglass shape. As a result, the trabecular area is relatively larger than in *L. africana* (n=3). In both species, the cortical thickness is greater on the medial side, although it is also locally thicker where it forms the deltoid tuberosity, which is thus entirely composed of compact bone (Fig. S13). Compact bone is naturally thin in the metaphyses, and extremely thin in the epiphyses. It is only slightly thicker, although much thinner than in the diaphysis, at the most proximal features of the proximal epiphysis (i.e. humeral head and greater trochanter), as well as in the olecranon fossa (Fig. 2A, B, D).

The medullary area is almost entirely filled with trabecular bone. At the GC, trabeculae are thicker and oriented from the thick medial cortex to the more distal, lateral epicondylar crest (Fig. 2C, F). The middle of the diaphysis above the GC is filled with thick trabeculae, oriented from the thick cortex under the deltoid tuberosity to the opposite side of the bone, slightly more distally (Fig. 2A, B, E). Distally to these thick trabeculae, the medullary area is free from trabecular bone for a few centimetres above the GC, forming a small medullary cavity. Thick trabeculae fill the center of the distal metaphysis, from the distal part of the diaphysis to the compact layer surrounding the olecranon fossa, delimiting the medial and lateral epicondyles (Fig. 2F). The trabeculae are much thinner in the epiphyses and the outer part of the metaphyses, where they are numerous and denser when closer to the cortex. In the proximal epiphysis, the trabeculae are denser in the medial and proximal parts of the humeral head; in the distal epiphysis, they are thick and dense on the lateral side (i.e. under the epicondylar crest) but thinner and less dense on the medial side. The centers of the epicondyles and trochlea are filled with much thinner and sparser trabeculae. The trabecular bone is highly anisotropic in the humerus. In the diaphysis, the trabeculae are highly anisotropic, oriented parasagittally (Fig. 2A, B). In the greater trochanter, the trabeculae are mainly oriented in the proximolateral direction (Fig. 2D). The trabeculae are less anisotropic in the center of the humeral head, but they grow more anisotropic medially, where they are oriented in the parasagittal axis. In the proximal metaphysis, the trabeculae are highly anisotropic, oriented parasagittally. In the distal metaphysis and epiphysis, the trabeculae are relatively thin and anisotropic, forming two proximally concave "bows" in the medial and lateral epicondyles respectively.

Despite some great similarities to adult specimens, the juvenile specimens also exhibit clear differences: the humerus is entirely filled with trabecular bone, leaving no medullary area free of bone; and the trabecular bone is overall less dense and the cortex is more symmetrical along the proximodistal axis than in adult specimens (Fig. 3A, B, C). Since the epiphyses are not fully formed, the cortex is virtually indistinguishable from the trabeculae in the bone extremities. Juvenile specimens also display much less anisotropic trabeculae: apart from the GC where they are clearly oriented along the parasagittal axis, the trabeculae are mostly isotropic. The youngest specimens of *L. africana* (n=10; one fetus, one neonate and several calves) display a marked asymmetry in the cortical thickness: the cortex appears relatively thicker on the lateral side than in adult specimens (Fig. 3D, E, F, G, H, I, Fig. S13). Although in older juveniles the cortex is relatively thicker on the medial side, this is also the case in the youngest specimens of *Elephas maximus* (n=5). The hourglass shape is visible even in the youngest specimens, centering on the GC, i.e. more distally than in adult specimens. Conversely, while juvenile specimens of *E. maximus* show a very clear hourglass-shaped organization of the cortex, it grows less marked in older juveniles and in adults (Fig. S7).

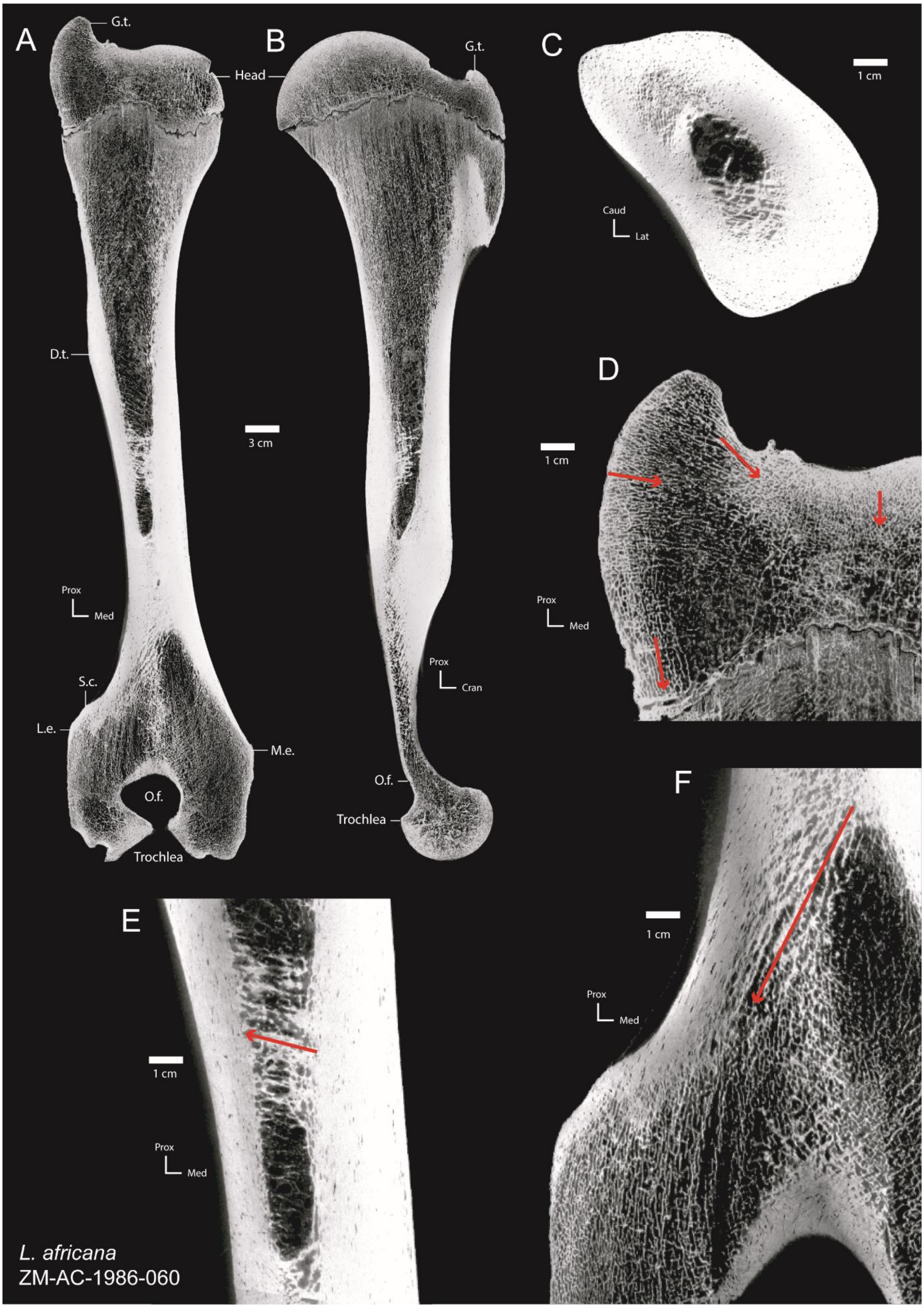

**Figure 2**. Virtual slices of the humeri of an adult (subadult) *Loxodonta africana* specimen (ZM-AC-1986-060) in (A, D, E, F), coronal, (B), sagittal, (C), and transversal views. The red arrows show the direction of trabeculae in highly anisotropic areas. D.t., deltoid crest, G.t., greater trochanter, L.e., lateral epicondyle, M.e., medial epicondyle, O.f., olecranon fossa, S.c., supracondylar crest. Caud, caudal, Lat, lateral, Med, medial, Prox, proximal.

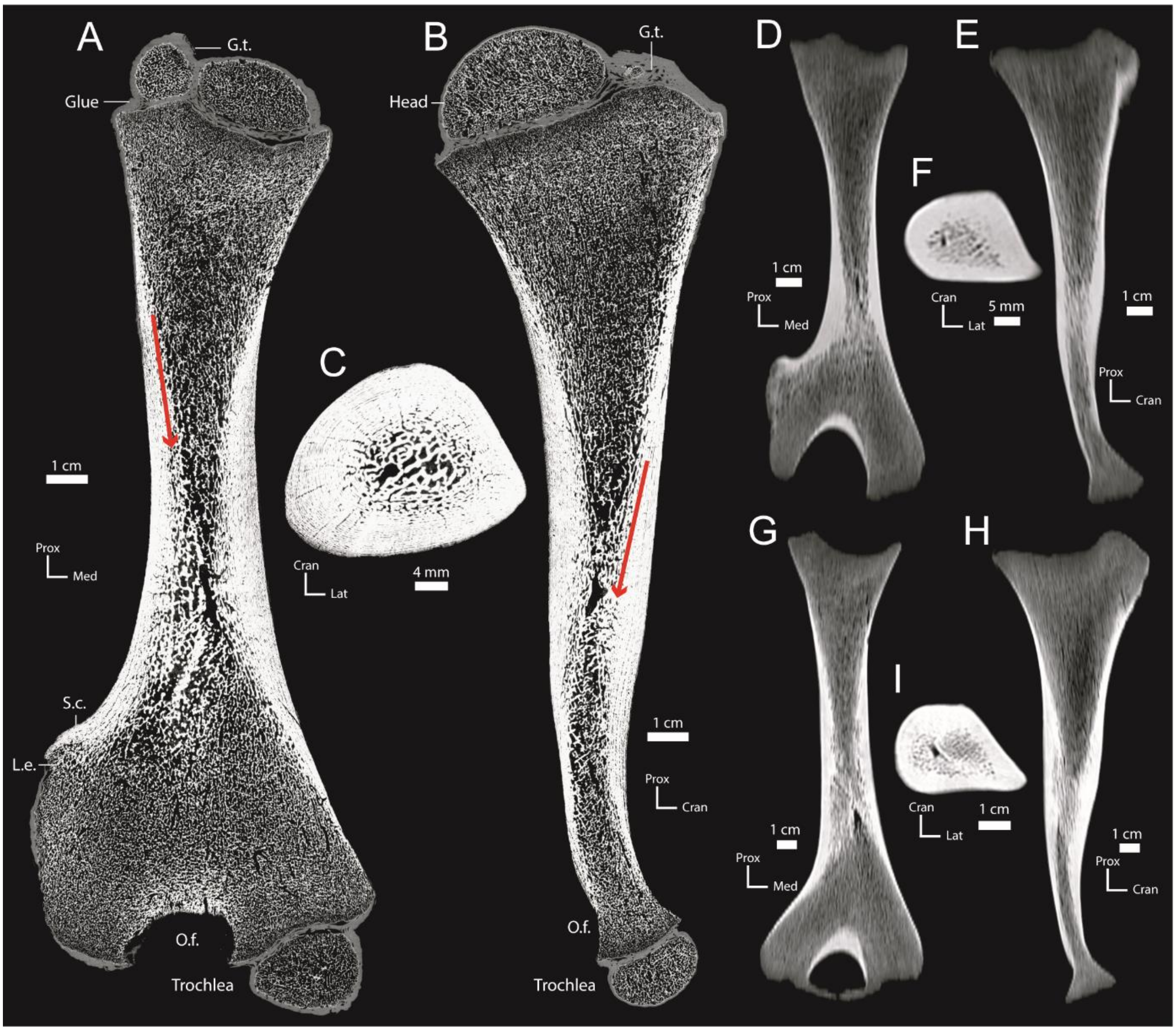

**Figure 3**. Virtual slices of the humeri of (A), (B), (C) a *Loxodonta africana* calf (ZM-AC-1961-69), (D), (E), (F) a *L. africana* fetus (NHMUK-1984.514) and (G), (H), (I) a *L. africana* neonate specimen (NHMUK-1962.7.6.8) in (A), (D), (G) coronal, (B), (E), (H) sagittal and (C), (F), (I) transversal views. The red arrows show the direction of trabeculae in highly anisotropic areas. D.t., deltoid crest, G.t., greater trochanter, L.e., lateral epicondyle, O.f., olecranon fossa, S.c., supracondylar crest. Caud, caudal, Lat, lateral, Med, medial, Prox, proximal.

### Radius

The radius of *E. maximus* and *L. africana* show a similar microanatomical organization. In adults (*E. maximus*, n=1; *L. africana*, n=2), the GC is located in the proximal half of the diaphysis (Fig. 4A, C). The cortex is relatively thick in the diaphysis and extremely thin in the epiphyses, especially in the distal one. In the proximal metaphysis, the cortical bone is thicker laterally (Fig. S13). In the proximal two-thirds of the diaphysis, the cortex is thicker medially, whereas in the distal third it is relatively thin and symmetrical mediolaterally, corresponding to the area where the radius's shaft crosses that of the ulna. The medullary area is entirely filled with trabecular bone, which is composed of relatively thick trabeculae around the GC (Fig. 4C, H) and of thinner and more numerous trabeculae in the rest of the diaphysis. The epiphyses are filled with thin trabecular bone made of thin struts and denser in the proximal than in the distal one. The boundary between the cortex and the medullary area is clear in the whole shaft. The trabeculae are overall isotropic in the proximal epiphysis and metaphysis, except at the most proximal extremity where they are orthogonal to the articular surface for the humerus (Fig. 4F). In the diaphysis and the distal epiphysis, the trabeculae are anisotropic, oriented parasagittally, except for the most caudal zone of contact with the ulna in which the trabeculae are slightly oriented along the mediolateral axis (Fig. 4B).

In juvenile specimens (*E. maximus*, n=3; *L. africana*, n=9), the GC is only slightly proximal to the middle of the diaphysis. The cortex is relatively thin; from the GC it becomes thinner both proximally and distally, forming an elongated hourglass-shaped medullary cavity (Fig. 4D, E). The cortex appears mostly symmetrical mediolaterally, apart from a slight thickening on the upper medial side of the shaft. In craniocaudal view, the cortex is thicker cranially. The compact bone is relatively thicker in the *L. africana* than in the *E. maximus* specimens. The medullary area is almost entirely filled with trabecular bone, except around the GC where it is partially filled with thick and sparse trabecular struts (Fig. 4I). The delimitation between the cortex and the medullary area is clear except in that area. The trabeculae are thick and sparse in the middle of the shaft and become thinner and denser towards the epiphyses, although the overall density of trabecular bone appears relatively low. The trabeculae are overall isotropic in the epiphyses and anisotropic in the diaphysis, where they are oriented parasagittally (Fig. 4D, E). The anisotropy is higher on the cranial side. Despite showing a much straighter diaphysis, the fetus specimen of *L. africana* displays no particular differences to the diaphysis of adults, and already displays the craniocaudal asymmetry of cortical thickness (Fig. S8).

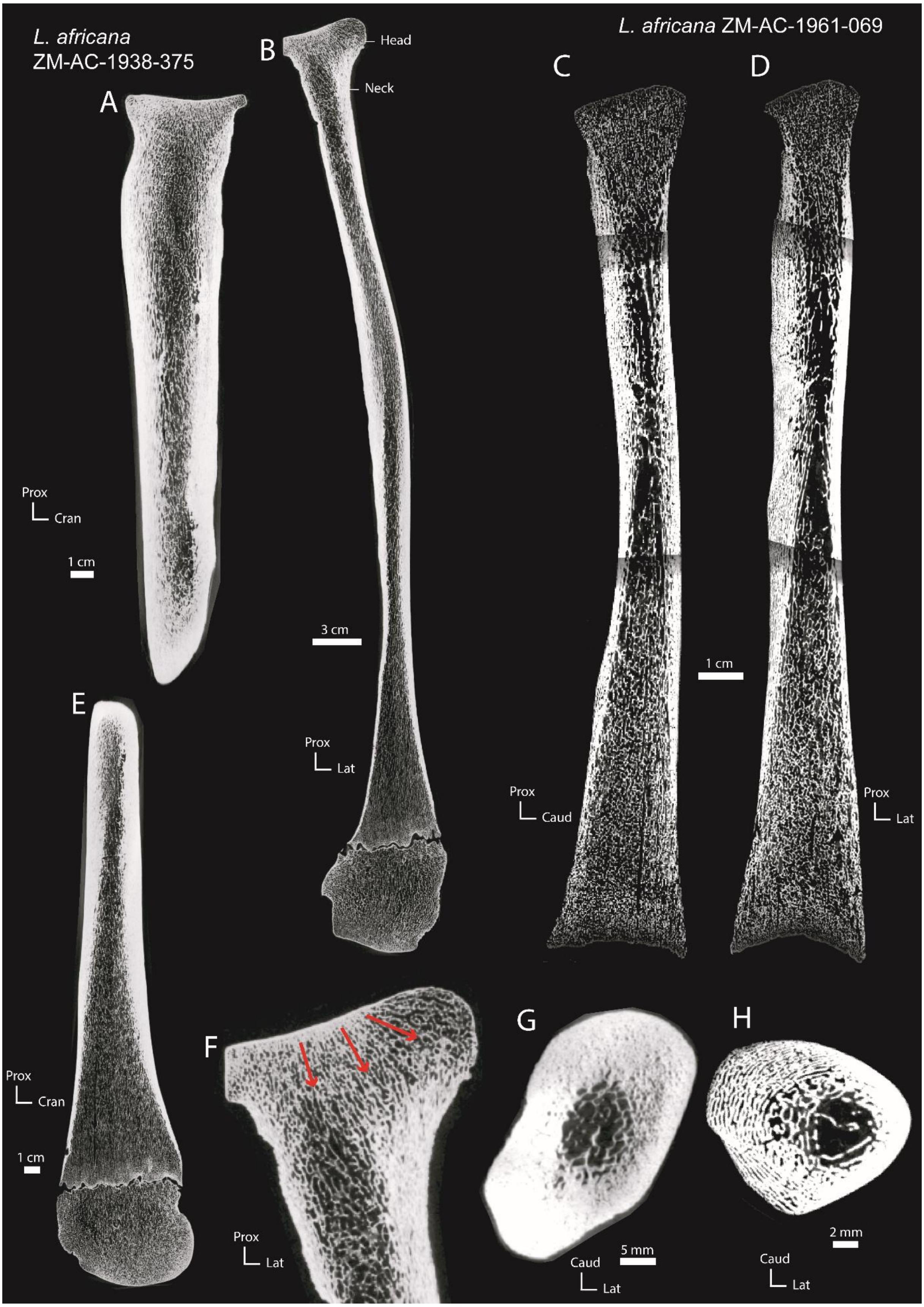

**Figure 4**. Virtual slices of the radius of *Loxodonta africana* specimens, (A), (B), (C), (F), (G) adult (subadult) (ZM-AC-1938-375) and (D), (E), (H), (I) calf (ZM-AC-1961-69) in (A), (D), (G), (H) coronal, (C), (E), (F), sagittal and (H), (I) transversal views. The red arrows show the direction of trabeculae in highly anisotropic areas. Caud, caudal, Lat, lateral, Med, medial, Prox, proximal.

## Ulna

In adult specimens of *L. africana* (n=1) and *E. maximus* (n=1), the GC is located in the upper part of the ulnar diaphysis, distal to the proximal metaphysis (Fig. 5A, B, C). The adult specimen of *L. africana* shows an asymmetrical cortex, thicker caudally in the proximal half of the shaft (Fig 5A, B). The cortex forms a clear hourglass shape, with two asymmetries: it is thicker medially along the whole diaphysis and thicker distally than proximally (Fig. S13). The adult specimen of *E. maximus* differs from the adult specimen of *L. africana* in cortical thickness: in *E. maximus* while the cortex is also asymmetrical in the diaphysis, being relatively thicker distomedially, this asymmetry is more pronounced in *E. maximus.* Additionally, because the cortex is relatively thinner in the proximal metaphysis of *E. maximus*, the hourglass shape is less marked than in *L. africana.* In both species, the cortex is much thinner in the epiphyses, with the notable exception of the trochlear notch; i.e. the articular surface for the humerus (Fig. 5E). In both species, the medullary area is almost entirely filled with trabecular bone, except for a few centimeters under the GC. The trabecular bone is made of thick and sparse struts around this area, and becomes denser with thin struts further away from the GC. In the proximal metaphysis and the proximal part of the shaft, the trabeculae are thin and isotropic in the center of the medullary area, whereas they are thicker and anisotropic close to the cortex, to which they are parallel. In the anconeus process, the trabeculae are oriented orthogonally to the articular surface for the humerus cranially; they are parallel to the cortex caudally; and these two anisotropies join each other in the center of the olecranon, forming an arch between the trochlear notch and the caudal part of the olecranon (Fig. 5E). In the most proximal part of the olecranon, the trabeculae are mostly isotropic laterally and anisotropic medially, where they are parallel to the cortex (Fig. 5F). The trabeculae are highly anisotropic and oriented along the parasagittal axis in the distal part of the diaphysis. In the proximal part of the diaphysis, the trabeculae are mostly isotropic, except proximally to the GC where they form an arch (Fig. 5I). Similarly, they are highly anisotropic in the center of the distal part of the distal epiphysis, where they are oriented orthogonally to the articular surfaces for the carpal bones (Fig. 5A); conversely, they are less anisotropic in the rest of the distal epiphysis.

In the juvenile *L. africana* specimens (n=10), the ulna's cortex shows the same hourglass shape as in the humerus and radius, although it is thicker in its proximal part (Fig. 5C, D). The medullary area is large and overall clearly delimited; as in the other bones the distinction is harder to make close to the GC due to the trabecular tissue being less dense and composed of thicker trabeculae (Fig. 5H). The craniocaudal and mediolateral asymmetries of the cortical thickness are already marked in fetal and neonate specimens. Juvenile specimens of *E. maximus* (n=4) show a clear hourglass-shaped organization of the cortical thickness, which is thicker on the medial side. The hourglass shape is less marked in older juveniles, and very slight in subadult specimens, resembling the distribution of adult specimens (Fig. S9).

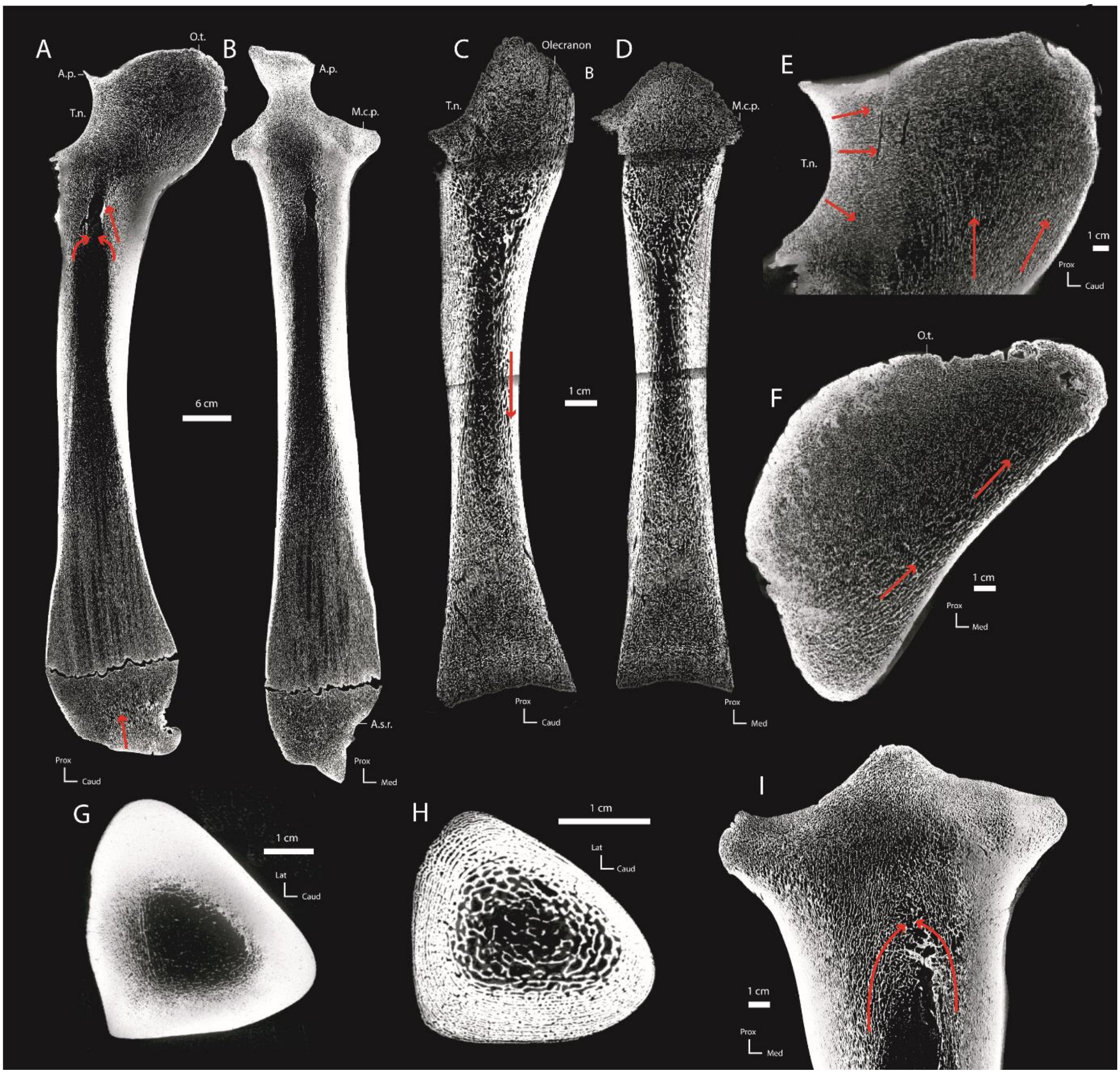

**Figure 5**. Virtual slices of the ulna *Loxodonta africana* specimens, (A), (B), (E), (F), (I) adult (subadult) (ZM-AC-1907-49), (C), (D), (G) juvenile (ZM-AC-1961-069) in (A), (C), (E), (I) coronal, (B), (D) sagittal and (F) (H), transversal views. The red arrows show the direction of trabeculae in highly anisotropic areas. A.p., anconeus process, A.s.r., articular surface for the radius, M.c.p., medial coronoid process, O.t., olecranon tuberosity, T.n., trochlear notch. Caud, caudal, Lat, lateral, Med, medial, Prox, proximal.

## Femur

In adult specimens of *E. maximus* (n=4) and *L. africana* (n=5), the GC is located in the proximal part of the femoral diaphysis (Fig. 6A, B, C, D) and the cortex is slightly thicker medioproximally and distolaterally. In the MNHN-ZM-AC-1997-30 specimen (*L. africana*), the cortex appears thickened in the femoral neck (Fig. 6A). Adult *L. africana* exhibit some variation in the symmetry of cortical thickness along the mediolateral axis, which does not appear to be linked to the size of the bone: for example, NHMUK-1961.8.9.82 is remarkably symmetrical, while specimen RVC-RSA shows a much thicker cortex on the medial side. Adult *E. maximus* specimens display a wider and more cylindrical medullary area than adult *L. africana* specimens; additionally, Asian elephants also display a thicker cortex under the lesser trochanter and in the distolateral part of diaphysis (Fig. 6C, D). Compact bone is extremely thin at the epiphyses, although there are several local thickenings: whereas it is much thinner than in the diaphysis, the cortex is thicker in the greater trochanter and, to a lesser extent, in the lesser trochanter than in the rest of the proximal epiphysis. The medullary area is entirely filled with trabecular bone (Fig. 6G, H). The trabeculae are thin and numerous; they are sparse around the GC and their network becomes progressively denser towards the proximal and distal epiphyses. In adults of both species, the trabeculae are overall isotropic in the femoral head; whereas they are highly anisotropic in the medial part of the neck, oriented parasagittally (Fig. 6A, C). The trabeculae are parallel to the surface of the bone in the rest of the proximal epiphysis, joining in an arch under the greater trochanter. The trabecular tissue is denser under the lesser trochanter, with parasagittally oriented trabecular struts; the trabeculae are slightly denser in the lesser trochanter. In the diaphysis, the trabeculae near the cortex are anisotropic and aligned along the parasagittal axis. In the center of the bone, however, they form arches of trabeculae oriented distally to proximally in the proximal half of the diaphysis and proximally to distally in the distal half, always from the outermost part of the medullary area to its core (Fig. 6A, B, C, D). In the distal epiphysis, the trabeculae are oriented parasagittally in the outermost parts of the medial and lateral epicondyles, and are more isotropic towards the center; the organization is similar in the condyles, although the anisotropy is more marked in *L. africana* than in *E. maximus* (Fig. 6A, C). The trochlea is filled with dense trabecular tissue made of isotropic trabeculae (Fig. 6B, D).

In juveniles (*E. maximus*, n=6; *L. africana*, n=10), the GC is located more distally, in the middle of the diaphysis (Fig. 6E, F). Unlike in the adult specimens, the cortex is asymmetrical, much thicker on the medial side (Fig. S13). The *L. africana* fetus shows a clear hourglass-shaped cortical distribution, which is also observable in the neonate specimen, although with a thickening on the caudal side (Fig. 6F). *E. maximus* calves and juveniles display an asymmetric cortex, thicker caudally (Fig. S10). However, in subadults the medullary area appears much larger with a more homogeneous cortex; the hourglass shape is thus less marked although still visible. Juvenile *L. africana* specimens show a similar anisotropic distribution to that of adult specimens; however, the trabecular bone is denser and composed of thicker trabeculae, although they become thinner in the extremities.

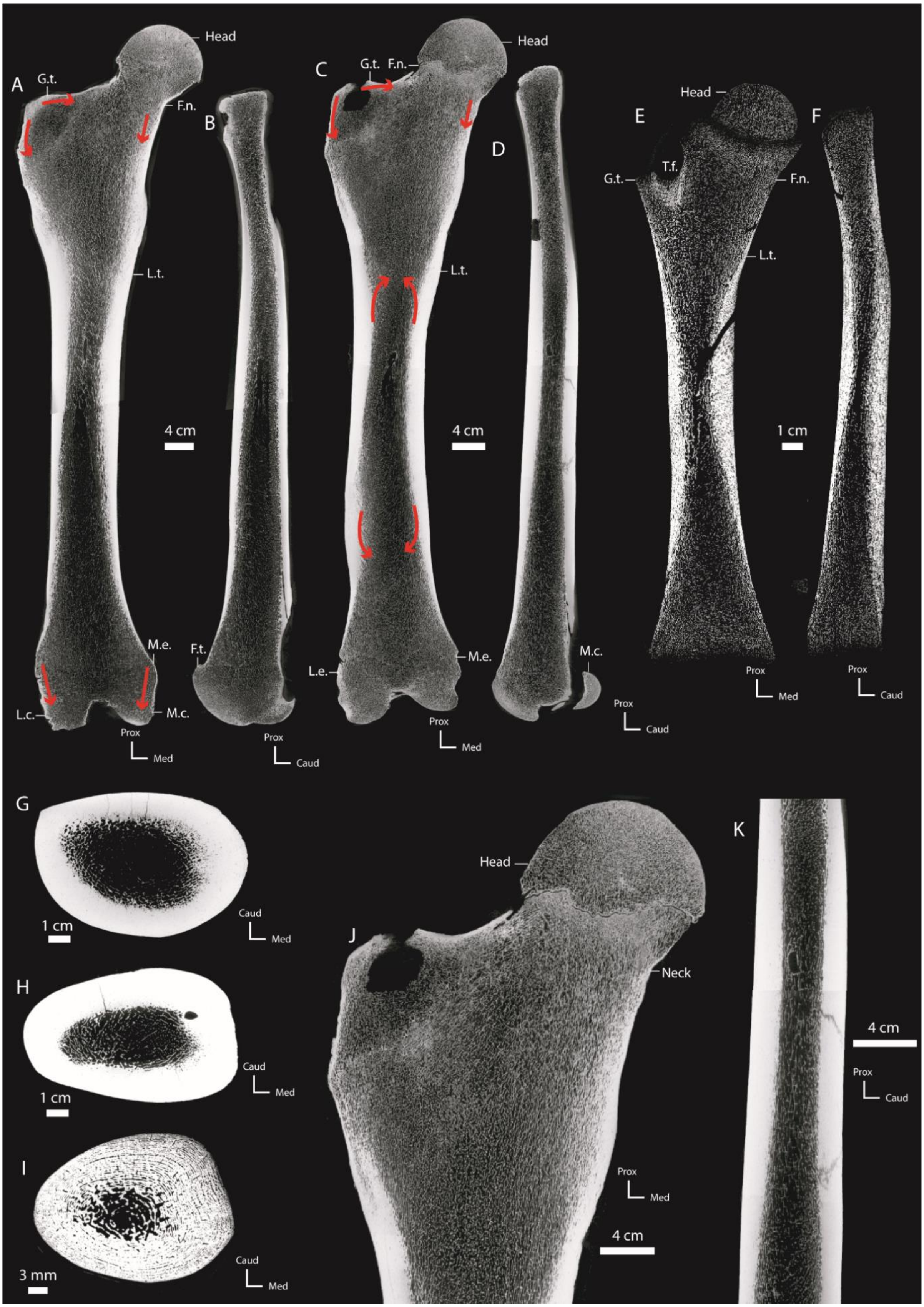

**Figure 6**. Virtual slices of the femur of (A), (B), (G), (J) an adult *Loxodonta africana* specimen (ZM-AC-1977-30G), (C), (D), (H), (K), an adult *Elephas maximus* specimen (ZM-AC-1883-1786), (E), (F), (I) a *Loxodonta africana* calf (ZM-AC-1961-069) in (A), (C), (E), (J) coronal, (B), (D), (F), (K) sagittal and (G), (H), (I) transversal views. F.t., femoral trochlea, F.n., femoral neck, G.t., greater trochanter, L.e., lateral epicondyle, L.t., lesser trochanter, M.c., medial condyle, M.e., medial epicondyle, T.f., trochanteric fossa. Caud, caudal, Lat, lateral, Med, medial, Prox, proximal.

## Tibia

In adult *E. maximus* specimens (n=3), the GC is located slightly distally to the middle of the shaft, and the medullary area is wider in the proximal part than in the distal part (Fig. 7A, B, K). The cortex is asymmetrical: it is thicker on the lateral side along the whole diaphysis, on the cranial side proximally, and on the caudal side distally. In the proximal epiphysis, the distinction between anisotropic trabeculae and compact bone is unclear; this area is either composed of thicker compact layer (although thinner than in the diaphysis), extremely dense trabeculae, or a mix of both. A similar organization is found in the distal articular surface (tibial cochlea) (Fig. 7J). The intercondylar eminence is covered with a thick layer of compact bone. The medullary area is entirely filled with trabecular bone. The trabeculae are thick around the GC, and get very thin and more densely packed in both metaphyses and epiphyses. The trabeculae are elongated and highly anisotropic in the entire bone, oriented parasagittally, except in the lateral epicondyle where they are oriented in the proximolateral direction (Fig. 7G). Although they display a similar overall organization, the adult specimens of *L. africana* (n=2) present several differences from adult *E. maximus*. Unlike in *E. maximus*, the cortex is much thicker on the lateral side of the diaphysis, and it is more symmetrical in the distal part of the diaphysis in the craniocaudal axis. The intercondylar eminence is filled with dense compact bone in *E. maximus* and with densely packed anisotropic trabeculae oriented orthogonally to the medial condyle in *L. africana* (Fig. 7C, D). The most striking difference is the presence of a medullary cavity: while in *L. africana* the cortical bone forms a more marked hourglass shape, the medullar cavity is not entirely filled with trabecular bone (Fig. 7C, D, L) as is observed in *E. maximus*; this medullary cavity extends from the proximal metaphysis to about a few centimeters above the distal metaphysis. While both species exhibit similar anisotropic orientations, in *L. africana* the trabeculae are not oriented parasagittally under the articular surface for the fibula, but rather orthogonally to that surface (Fig. 7I).

The microanatomy of juvenile *L. africana* specimens (n=9) is overall similar to that of adults of the same species. Juvenile *E. maximus* (n=5) show a more distal GC than adults, and a much thicker cortex in the craniodistal part of the shaft, not under the crest (Fig. 7F, Fig. S13). The asymmetry of the cortex is more marked in coronal view (Fig. 7E, F, M); the hourglass shape of the medullary area is also clearer, especially in the distal half of the bone where it is relatively larger than in adult and subadult specimens (Fig. S11). The trabeculae are less elongated, and are less anisotropic in the bone extremities, although the anisotropy is clear closer to the cortex and to the GC. The subadult specimen of *E. maximus* displays a relatively larger medullar cavity, and the asymmetry of the cortex is more marked on the medial and cranial sides. The trabecular tissue is relatively dense at the GC, although the trabeculae are much thinner than in adult specimens, and with a lower anisotropy. In the proximal epiphysis the tibial crest is characterized by marked anisotropy. (Fig. 7H).

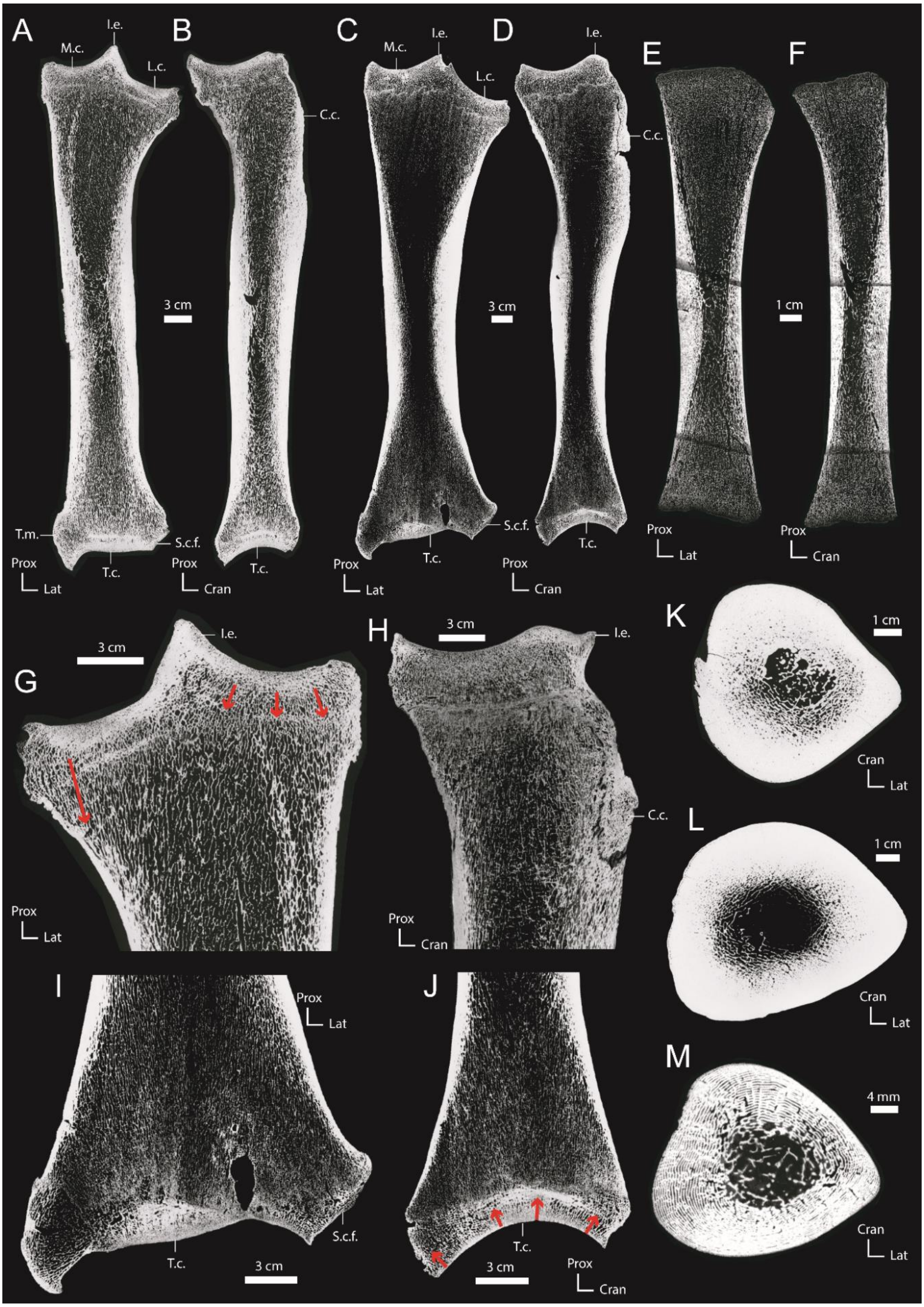

**Figure 7**. Virtual slices of the tibia of (A), (B), (G), (K) an adult *Elephas maximus* specimen (ZM-AC-1983-082), (C), (D), (I), (J), (L) an adult *Loxodonta africana* specimen (ZM-AC-1907-49), (E), (F), (M) a *Loxodonta africana* calf (ZM-AC-1961-069), (H) a subadult *Elephas maximus* specimen (ZM-AC-1936-280), in (A), (C), (E), (G), (I) coronal, (B), (D), (F), (H), (J) sagittal and (K), (L), (M), transversal views. C.c., cranial crest, I.e., intercondylar eminence, L.c., lateral condyle, M.c., medial condyle, S.c.f., surface of contact with the fibula, T.c., tibial cochlea, T.m., tibial malleolus. Caud, caudal, Lat, lateral, Med, medial, Prox, proximal.

## Fibula

In adult *E. maximus* specimens (n=3), the GC is located in the distal part of the diaphysis (Fig. 8A, B, C). The cortex is thick, surrounding a thin medullary area. The cortex is remarkably consistent in thickness along the shaft, thus not forming the hourglass shape observed in the other five bones. The medullary area is almost entirely filled with trabecular bone, except in the more central region in the proximal part of the diaphysis, where the medullary area is relatively the largest (Fig. 8A, B, C, J). The trabeculae are thin and numerous, particularly in the medullary area where they are highly anisotropic and oriented parallelly to the shaft. The trabeculae are anisotropic in the distal half of the shaft and in the distal epiphysis where they are oriented parasagittally in the lateral part (Fig. 8L, M). In the proximal epiphysis, the anisotropy is expressed by forming arches from the caudal to the cranial sides in the proximal metaphysis (Fig. 8B, J, K); proximally, the head of the fibula is filled by isotropic trabeculae (Fig. 8A, B). The fibulae of adult *L. africana* specimens (n=2) are very similar to that of the adult *E. maximus*, with two exceptions. First, the cortical thickness is highly asymmetrical in the mediolateral axis, with a much thicker cortex on the lateral side of the diaphysis (Fig. 8E, O, Fig. S13). Second, the trabeculae display very low anisotropy, only slightly oriented parasagittally in the diaphysis and the metaphyses (Fig. 8E, F, G). The trabeculae are however highly anisotropic in the articular surface for the tibia, oriented orthogonally to the articular surface of the malleolus (Fig. 8M), while in *E. maximus* the anisotropy is much more marked in the articular surface for the carpus (Fig. 8L). The subadult specimen of *E. maximus* displays a similar microanatomical distribution, apart from a more proximal location of the GC and an overall lesser degree of anisotropy (Fig. 8H, I, J, P). Similarly, juvenile specimens of *L. africana* (n=6) resemble adult specimens (Fig. S12).

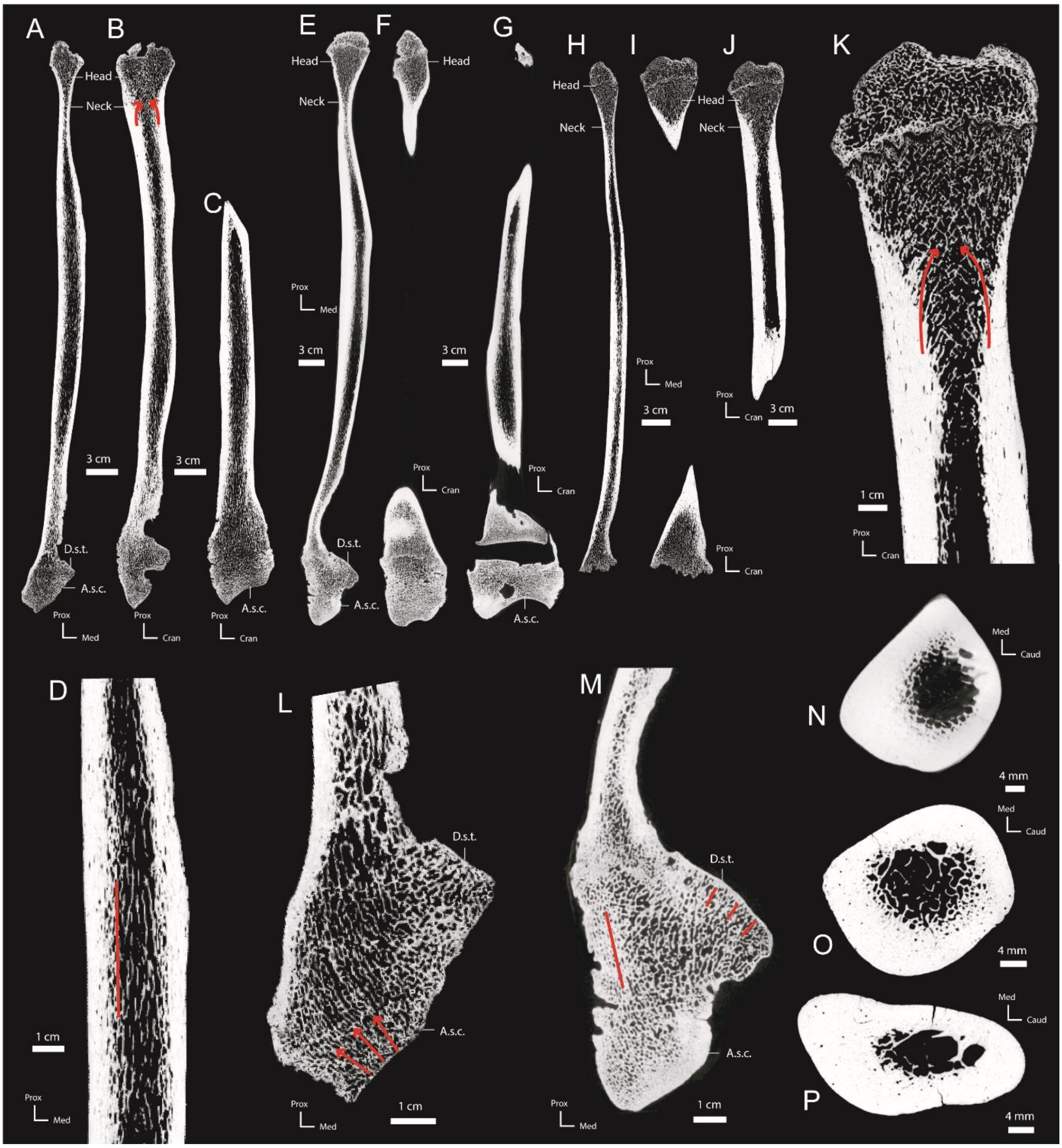

**Figure 8**. Virtual slices of the fibula of (A), (B), (C), (D), (L), (N) an adult *Elephas maximus* specimen (ZM-AC-1983-282), (E), (F), (G), (M), (O), an adult *Loxodonta africana* specimen (ZM-AC-1907-49), (H), (I), (J), (K), (P) a subadult *Elephas maximus* specimen (ZM-AC-1936-280), in (A), (D), (E), (H), (L), (M) coronal, (B), (C), (F), (G), (I), (J), (K) sagittal and (N), (O), (P) transversal views. A.s.c., articular surface for the carpus, D.s.t., distal surface of contact with the tibia. Cran, cranial, Med, medial, Prox, proximal.

## Size and robustness analyses

There is no significant difference of robustness between the long bones of *E. maximus* and *L. africana*.

Depending on the bone considered, we observed variation of robustness during ontogeny (Fig. 9, Table 2, 3). In the humerus and tibia of both species and in the ulna of *L. africana*, robustness overall appears to remain stable during ontogeny. Robustness decreases during ontogeny in the ulna of *E. maximus* (p=0.01; $r^2$=0.53). Conversely, the robustness of the femur appears to increase with ontogenetic stage in both species, although it is significant in *L. africana* only (*E. maximus*: p=0.04; $r^2$=0.16, *L. africana*: p=0.01; $r^2$=0.44). The fibula provided mixed results, with *E. maximus* displaying a relatively stable robustness during ontogeny, whereas *L. africana* displays a higher robustness in juvenile specimens than in calves and adults. During ontogeny, the humerus and femur display opposite trends: the humerus is slightly more robust in *L. africana* than in *E.* maximus calves, and less robust in adult *L. africana* than in *E. maximus* specimens; it is the opposite in the femur. Although the radius and ulna show different robustness levels in juveniles of both species (the radius is more robust in *L. africana*, the ulna is more robust in *E. maximus*); their robustness is consistent between the two species in both calves and adult specimens. Similarly, the tibia and fibula appear to be consistently more robust in adult specimens of *E. maximus* than in adult specimens of *L. africana.*

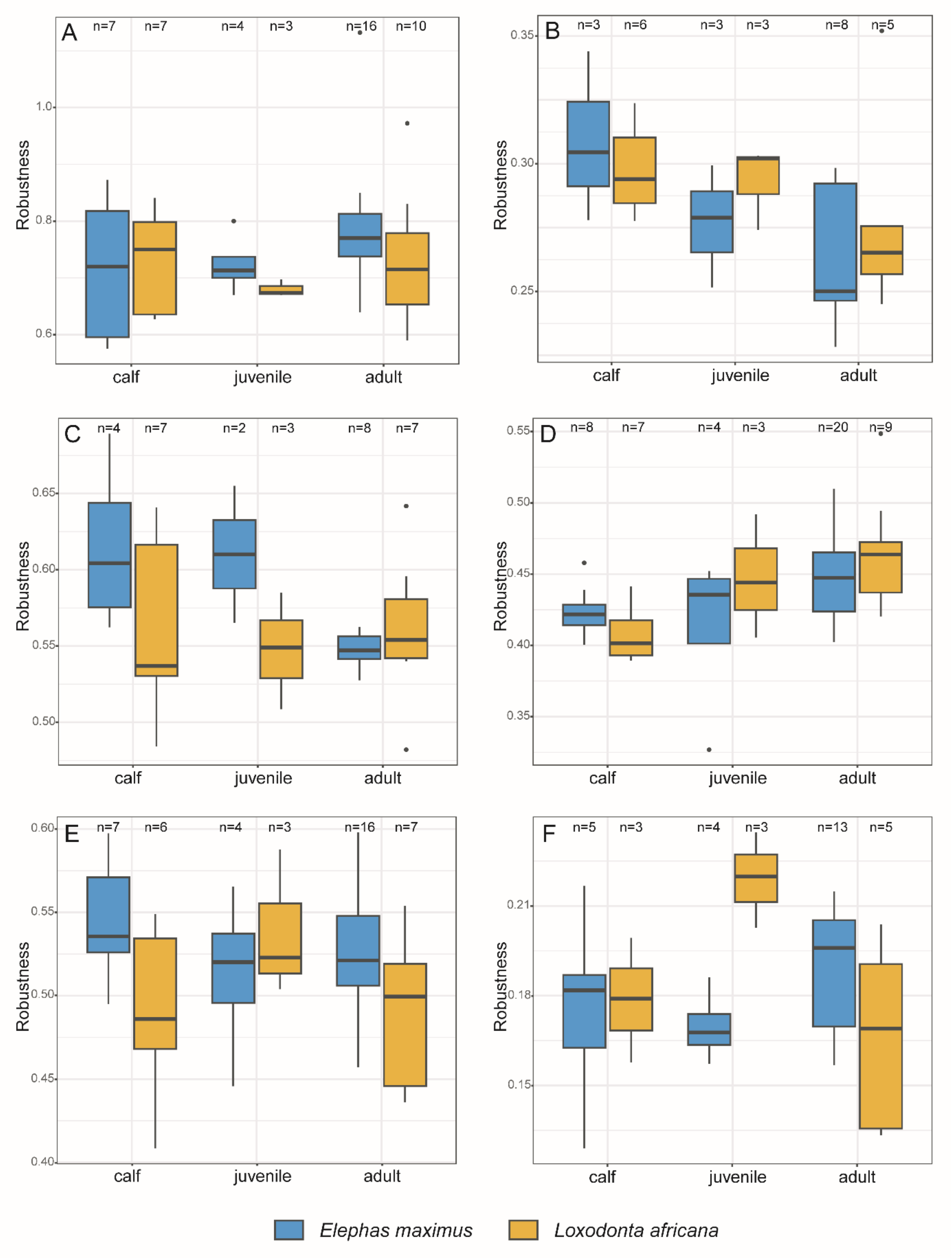

**Figure 9**. Boxplot of the robustness values for the three age (ontogenetic stage) groups (calf, juvenile, adult) in *E. maximus* and *L. africana*. (A), humerus, (B), radius, (C), ulna, (D), femur, (E), tibia, (F), fibula.

**Table 2**. FDR corrected results of the ANOVAs testing for robustness variation between *Elephas maximus* and *Loxodonta africana* depending on the ontogenetic stage of the individuals. *p*, p-value; $r^2$, determination coefficient value. Significant results are in bold.

| **Bone** | **Calf** | **Juvenile** | **Adult** |
|---|---|---|---|
| **Humerus** | n=14 | n=7 | n=26 |
| | p=0.95 | p=0.47 | p=0.47 |
| | $r^2$<0.01 | $r^2$=0.26 | $r^2$=0.06 |
| **Radius** | n=9 | n=7 | n=12 |
| | p=0.67 | p=0.60 | p=0.60 |
| | $r^2$=0.06 | $r^2$=0.19 | $r^2$=0.06 |
| **Ulna** | n=11 | n=5 | n=14 |
| | p=0.47 | p=0.56 | p=0.60 |
| | $r^2$=0.17 | $r^2$=0.40 | $r^2$=0.05 |
| **Femur** | n=15 | n=7 | n=29 |
| | p=0.47 | p=0.60 | p=0.56 |
| | $r^2$=0.16 | $r^2$=0.12 | $r^2$=0.03 |
| **Tibia** | n=14 | n=7 | n=24 |
| | p=0.22 | p=0.65 | p=0.25 |
| | $r^2$=0.31 | $r^2$=0.09 | $r^2$=0.15 |
| **Fibula** | n=8 | n=7 | n=21 |
| | p=0.95 | p=0.10 | p=0.45 |
| | $r^2$<0.01 | $r^2$=0.81 | $r^2$=0.16 |

**Table 3**. FDR corrected results of the t-tests testing for robustness variation between calves and adult specimens of *Elephas maximus* and *Loxodonta africana*. *p*, p-value; $r^2$, determination coefficient value. Significant results are in bold.

| **Bone** | ***Elephas maximus*** | ***Loxodonta africana*** |
|---|---|---|
| **Humerus** | n=27 | n=20 |
| | p=0.47 | p=0.95 |
| | $r^2$=0.09 | $r^2$<0.01 |
| **Radius** | n=14 | n=14 |
| | p=0.24 | p=0.60 |
| | $r^2$=0.37 | $r^2$=0.10 |
| **Ulna** | n=14 | n=16 |
| | **p=0.01** | p=0.95 |
| | **$r^2$=0.53** | $r^2$<0.01 |
| **Femur** | n=32 | n=19 |
| | p=0.24 | **p=0.01** |
| | $r^2$=0.16 | **$r^2$=0.44** |
| **Tibia** | n=29 | n=16 |
| | p=0.95 | p=0.24 |
| | $r^2$=0.07 | $r^2$<0.01 |
| **Fibula** | n=22 | n=14 |
| | p=0.56 | p=0.69 |
| | $r^2$=0.06 | $r^2$=0.05 |

# DISCUSSION

## General pattern and weight bearing adaptations

Our study indicates that the six long bones of extant elephant limbs share global microanatomical patterns to accommodate weight-bearing: *E. maximus* and *L. africana* show bones with a relatively thick cortex and otherwise almost entirely filled with trabecular bone, as is expected in graviportal animals (Houssaye *et al.*, 2016; Lefebvre *et al.*, 2023).

The two species exhibit a peculiar distribution of cortical bone, which is particularly thick around the growth center, forming two "cones" or an "hourglass-shaped" distribution. In the absence of secondary bone deposits in the medullary area, such cortical bone thickness is only possible near the growth center, where compact bone is first deposited during development. Thicker cortex allows for better resistance to load compression. This thicker cortex could also correspond to the area where the forces are the greatest when bones are under loading, thus maintaining stresses at safe magnitudes. A similar pattern is observed in extant rhinoceroses (Etienne, 2023) and hippos (Houssaye *et al.*, 2021), where the higher anisotropy of trabecular bone surrounding the GC indicates potential for high stresses, suggesting that in those taxa bone resorption is limited in order to maintain a thick layer of cortical bone able to withstand massive loads. Interestingly, whereas some elephant-sized sauropods show a cortical thickening near their presumed GC, it is much less marked than in hippos and rhinos (Lefebvre *et al.*, 2023), indicating that other anatomical features are involved in weight-bearing, thereby releasing the constraints on bone microanatomy (Lefebvre *et al.*, 2023). In elephants, the hourglass shape is more marked in the forelimb than in the hindlimb. This result is consistent with the higher mechanical loads placed on the forelimb than on the hindlimb (Lessertisseur & Saban, 1967; Hildebrand, 1982; Polly, 2007), suggesting varying bone resorption depending on the bone considered and the mechanical constraints involved.

If a thicker cortex indicates resistance to a mechanical load (e.g., Currey, 2002), we can consider cortical thickness (like bone robustness and articular surfaces) as a proxy for load distribution and transmission along the bone; an asymmetrical cortical thickness along the shaft indicates an asymmetrical load on the bone. With the exception of the fibula, the cortex is asymmetrical in all long bones, with the medial side of the shaft displaying a slightly thicker cortex than the lateral side in the humerus, radius and ulna, and the lateral side displaying a thicker cortex in the femur and tibia. This unequal distribution suggests that in elephants, the mechanical load is placed more medially in the forelimb, which is partly inconsistent with the findings of Panagiotopoulou et al. (2012, 2016) who observed a general trend of lateral loading in the dynamic foot pressure of Asian and African savanna elephants. This discrepancy between the cortical thickness of the limb long bones and the foot pressures might be explained by a more medially angled ground reaction force, by a redistribution of the mechanical load in the autopod, by the footpad and "predigits" supporting the autopods (Hutchinson *et al.*, 2011), or a combined effect. This mediolateral thickness distribution was not observed in columnar sauropods (Lefebvre *et al.*, 2023) but was observed in the humerus of rhinos and hippos (A. Houssaye, pers. obs.; (Houssaye *et al.*, 2021), suggesting a similar trend of body mass support in mammals via modification of the regions of the bones closer to the center of mass.

Trabecular bone is more compliant than compact bone (Currey, 2002). Although we observed thickenings of the compact layer in the epiphyses in certain bones (e.g., greater trochanter of the humerus), it always remains far thinner than in the diaphysis, allowing some degree of compliance in the bone extremities. Consistently with Nganvongpanit et al. (2017), we observed that all six bones are filled with trabecular bone, leaving almost no medullary cavity. This kind of organization is typical of large quadrupeds (Houssaye *et al.*, 2016), in which the increased trabecular volume appears to improve load distribution along the bone and thereby capacity for weight support (Currey, 2002). Additionally, the trabeculae are highly anisotropic: in the diaphysis of all bones, they are oriented parasagittally, thus maximizing axial load distribution. This reflects the immense mechanical load and the angle of application, corresponding to the columnar stance of elephants, in which the long axis of their limbs is nearly orthogonal to the ground during standing and slow walking (e.g., Ren et al., 2010). These results are consistent with those of Lefebvre et al. (2023) who described a similarly filled medullary area in sauropods, linked to the heightened mechanical load placed on their columnar limbs.

The six long bones of extant elephants thus share a global microanatomical pattern that is adapted to deal with heavy weight-bearing. However, these bones do not participate equally in the movement and support of the body (Lessertisseur & Saban, 1967; Hildebrand, 1982; Polly, 2007; Ren *et al.*, 2010; Etienne *et al.*, 2021). We present evidence here that the different roles of the bones are reflected in their microanatomy.

## Forelimb

Although the overall inner organization of the bony tissue appears to be linked with weight support, the humerus also displays a particularly thick cortex in the deltoid tuberosity, which bears the deltoid muscle, a strong shoulder abductor and flexor (Shindo & Mori, 1956a). Additionally, the trabeculae are highly anisotropic in the greater trochanter, reflecting the large muscular forces applied by shoulder adductors and stabilisers. These results are in accordance with the external morphology of the humerus, which highlights the variety of forces it is subjected to, due to its place closer to the thorax and its involvement with the shoulder joint.

Unsurprisingly, the radius shows only limited adaptations of its microanatomy, presumably reflecting its limited involvement in weight-bearing. This is coherent with its relatively reduced width and robustness compared to the ulna, which is the main zeugopodial weight-bearer in the forelimb (Smuts & Bezuidenhout, 1993; Christiansen, 1999). Interestingly, despite its reduced participation in load-bearing the radius displays a thicker cortex and denser trabeculae in the proximal epiphysis than in the distal one, suggesting a link to its contact with the humeral trochlea. This bony distribution might reflect a more lateral loading in the zeugopod, although it may also be linked to the relatively small size of the radial proximal epiphysis compared to the distal one; the thick compact bone and dense trabecular bone might compensate for the relatively small articular surface.

The microanatomy of the ulna appears to be mostly adapted to weight-bearing, with few adaptations linked to muscular forces except in the olecranon. This result is coherent with the role of the ulna in elephants: unlike in other large-bodied taxa (e.g. rhinos, Etienne, 2023; bovids, Barone, 2020; equids, Barone 2020), the ulna is the main bearer of the weight in the forelimb's zeugopod, forming a "pillar" bearing relatively weak muscles (Smuts & Bezuidenhout, 1993; Larramendi, 2015). It is thus unsurprising to observe a microanatomical distribution mainly dedicated to weight-bearing. Due to the presence of the olecranon tuberosity, which is extremely developed caudally in elephants, the ulna presents a caudal curve. As a result, the ulna is not fully aligned orthogonally to the ground, and thus to the axial compressive load. Such a shape might be at greater risk of failure under the presumably large compression forces without compensation of the microanatomical organization. Accordingly, the cortex is much thicker on the caudal side of the ulna, especially in the proximal two-thirds; i.e. from the olecranon to below the mid-diaphysis. This thick cortex should allow the support of the mechanical load despite the bone curvature. In addition, the thicker cortex medially corresponds to the area of contact with the radius, implying an adaptation to load transfers between the two bones. The trabeculae are highly anisotropic in the medial part of the olecranon, consistently with its role as a mechanical lever and insertion of the *triceps brachii* muscle, a strong elbow extensor. Deeper inside the olecranon, the loads of this muscle appear to be distributed via an arch of trabecular bone, transmitting them to the rest of the ulna.

The microanatomy of the forelimb bones thus reflects the different constraints to which the bones are subjected: whereas the three bones are adapted to carry a heavy weight, the inner organization of the humerus appears adapted to heightened and more diverse muscular demands compared to the radius and ulna, with local thickenings of the cortex and areas of high anisotropy corresponding to muscular insertions. The contact between the radius and ulna is reflected in the microanatomy of their distal epiphyses: the distal radial epiphysis shows a slight trabecular anisotropy oriented orthogonally to the area of contact with the ulna; whereas the distal ulnar epiphysis shows slightly denser trabecular bone. However, these local adaptations are relatively modest as compared to the overall microanatomical organization of the two bones: the ulnar microanatomy reflects its role as main weight-bearer in the zeugopod, whereas the radius shows limited adaptation of its microanatomy corresponding to its lesser role in weight-bearing.

### Hindlimb

The hindlimb displays an opposite pattern of reaction to muscular loading in its limb segments compared to the forelimb: except for the anisotropy of the greater trochanter and the slightly thicker cortex of the lesser trochanter in Asian elephants, the femur shows no clear adaptation to muscular loads. The femur instead displays a cortical thickness distribution and anisotropy adapted to vertical loading, with isotropic trabeculae in areas subjected to what may be highly variable forces (i.e. femoral head and trochlea). The tibia, however, shows adaptations linked to muscular loads: the tibial crest in particular is reinforced by thick compact bone, supporting large forces from the *quadriceps femoris* muscles, which are powerful knee extensors. In addition, the tibia exhibits a thicker cortex in the caudal part of the distal two-thirds of the diaphysis. This distribution is reminiscent of what is observed in the ulna, i.e. a cortical thickening counter-balancing the curvature of the bone.

Although the fibula is often reduced or absent in many animals (Barone, 2020), it is present with a diaphysis in graviportal species. Like in rhinos (Etienne, 2023) and sauropods (Lefebvre *et al.*, 2023), in elephants the fibula is the only long bone that does not exhibit the hourglass-shaped organization of cortical thickness, instead possessing a relatively thinner cortex compared to other bones. Despite lacking the typical weight-bearing adaptations of other long bones, the fibula is relatively much larger than in other large-bodied taxa such as hippo and rhinos, in which it is greatly reduced compared to the tibia. In elephants, the fibula features anisotropic trabeculae in the diaphysis and especially in the distal epiphysis, suggesting an adaptation to withstand considerable loads. Nonetheless, the microanatomical distribution represents a convergent adaptation among extant large-bodied taxa, highlighting the fibula's limited role in weight support compared to the other long bones. This convergence suggests distribution of roles within the hindlimb's zeugopod, with the tibia bearing most of the weight and the fibula primarily acting as an ankle stabilizer (Etienne, 2023).

## Interspecific variation

While *L. africana* and *E. maximus* share similar microanatomical patterns of the limb bones that are remarkably adapted to weight-bearing, there are some notable differences between the two species. In both species, the long bones (except the fibula) display an hourglass-shaped distribution of the cortical thickness, more marked in *L. africana* than in *E. maximus.* This growth cone is less resorbed in elephants than in most taxa (Houssaye *et al.*, 2021), resulting in this hourglass distribution, well suited to deal with axial compression. The relative difference in thicknesses could be explained by the greater body mass of *L. africana*. The two species exhibit important size and mass variations depending on age and sex (Wittemyer, 2011), so that we would expect the cortical thickness distribution to reflect body mass and thereby sex as well. Despite the general lack of information on the sex of our specimens, this pattern is evident at different ontogenetic stages and in both males and females, suggesting that this pattern is species-specific rather than linked to mass only. Bader et al. (2023) stated that Asian elephants showed overall more robust bones than African savanna elephants, and linked this morphological variation to habitat difference between the two species. *E. maximus* might rely on a relatively more complex muscular support than *L. africana*, possibly due to an increased demand for stabilization and dexterity to navigate the soft and humid soils of the closed forest, as opposed to the relatively less challenging environment of the open savanna and its stiff, dry substrate. The difference in cortical thickness distribution observed here might thus be linked to difference in contact forces and/or varying muscular demands, although the complexity of factors involved calls for further experimentation on the impact of substrate on loading distribution and muscular stresses. Regardless of the effect of environment, the cortical thickness distribution in *L. africana* appears to counterbalance thinner morphology of the bones, allowing for greater mechanical loading for a slimmer external morphology.

Our results also indicate that the larger diaphyses of the limb long bones in *E. maximus* are not associated with a thicker cortex, but with a larger medullary area. Most interestingly, while the bones are filled by trabecular bone in both *E. maximus* and *L. africana*, the tibia of *L. africana* exhibits a medullary area that is not entirely filled with trabecular bone. This suggests a relaxation of constraints on the microanatomy, which are accommodated by the bone's external morphology.

Interestingly, we found no difference in long bone robustness between the adult specimens of *E. maximus* and *L. africana* in our sample, unlike what was observed in Bader et al. (2023), who noted significantly more robust humeri, ulnae and tibiae in *E. maximus* than in *L. africana*. This discrepancy between the results despite a largely overlapping sample is probably due to the maximal length measurements of the bones, which were measured on the whole bone in Bader et al. (2023) and only on the diaphysis in the present study. These results are consistent with those of Herridge (2010) who performed robustness analyses using the diaphyseal length on elephant limb bones. This suggests that the epiphyses add enough length to the bone to change the robustness calculation and thus the overall variation in bone robustness between the two species: the epiphyses are proportionally longer in *L. africana*, so that this species displays less robust bones when including the epiphyses in the length measurements. Our observations here underline the importance of using the whole bone length into account when comparing robustness in adult specimens. While we describe here a number of interspecific variations, our findings may be limited by the relatively small sample size, which could affect the robustness of our conclusions regarding species-specific patterns in bone microanatomy. While the observed patterns offer valuable insights, further studies with larger sample sizes would be necessary to confirm test these trends and draw more definitive conclusions.

## Ontogenetic patterns of microanatomical variation

Bones are subjected to numerous constraints during growth. Changes in locomotor behaviour and mass increase, in particular, greatly modify the mechanical loads placed on the appendicular skeleton. Our results indicate an ontogenetic change in the trabecular organization of the six long bones: juvenile specimens exhibit a lower trabecular density and lesser degrees of anisotropy than adult specimens do. This result is consistent with process known as bone functional adaptation (Ruff *et al.*, 2006), in which the internal structure of the bone adapts in response to loading, thus reflecting the intensity and orientation of the forces applied on the bones. Adult elephants weigh between 30 to 90 times their birth weight (Wittemyer, 2011). In juveniles, body mass is proportionally much lower: as body mass scales up much faster than body length (Schmidt-Nielsen, 1984), the ratio of mass to linear size is much lower in calves, thus not necessitating the high density and anisotropy levels observed in adults. However, the shift in trabecular microanatomy happens relatively early in the life of the animals, as older juveniles and subadult specimens display much denser and anisotropic trabeculae, highlighting how quickly the bones are able to model their geometry (*sensu* Frost, 1982) and adapt to changing constraints.

Bones of the forelimb display a more symmetrical distribution of the cortical thickness in juveniles than in adults, in which the cortex is slightly thicker on the medial side of the diaphysis, suggesting a greater mechanical load medially. This change of cortical thickness distribution suggests either that the smaller body mass in juveniles is not enough to warrant an asymmetry of the cortical thickness, or that loads are more evenly distributed along the mediolateral axis in juveniles than in adult specimens. Additionally, this shift is evident in bones of the forelimb only, suggesting a link with the proportionally greater load placed on the forelimb, possibly correlated with the weight of the tusks, which except in the tuskless females of *E. maximus* can add between 10 and 100 kg (Shoshani & Tassy, 2005) in adult elephant specimens.

Our observations indicate that the hourglass-shaped distribution of the cortical thickness is already present in juvenile specimens, even before birth. This observation indicates an inhibition of the cortical bone resorption early during growth, and thus underlining the presence of this microanatomical organization for weight support even in the earliest stages of life. While this distribution is visible in both *E. maximus* and *L. africana*, it is less marked in adult *E. maximus* specimens: the humerus, ulna and femur of calves and juveniles show a clear hourglass-shaped distribution, while it is less marked in subadult and adults. Conversely, in *L. africana* the hourglass shape is maintained and just as visible in adult specimens, reflecting the interspecific difference of strategy for weight-bearing.

Interestingly, the radius of both species shows that the cortical distribution is similar in juveniles and in adults, while the anisotropy greatly varies during growth. In all other bones, juveniles display differences in cortical thickness distribution compared to adults, although proportionally less marked than the anisotropy variation. This supports the overall observation that trabecular density and anisotropy levels appear to change later and to a higher degree than cortical bone thickness during the growth of the individual (Hildebrand, 1982; Currey, 2002). Finally, juvenile specimens show little to no cortical thickening linked to areas of muscular insertions. Cortical thickening at insertions is particularly well exemplified in the tibia, where the cranial crest gets progressively thicker with age, responding to the growing muscular forces applied by the quadriceps muscle group. These observations again reflect the bone functional adaptation during growth.

The humerus and tibia display a similar robustness throughout ontogeny, indicating isometric growth in both species. The more homogeneous cortical thickness observed in the humerus of adult *E. maximus* specimens is thus not associated with robustness variation, indicating that changes in the microanatomy are not always associated with the outer shape of the bone. The growth pattern of the ulna appears to differ between the two species: in *L. africana*, the ulna displays similar robustness values during growth, whereas in *E. maximus* the ulna becomes less robust. However, our sample of adult ulnae is extremely limited, so further analyses are needed to resolve this matter.

In both species, the radius is relatively less robust in adults than in juvenile specimens, following the general rule of robustness decrease during growth in quadrupedal mammals (Carrier, 1996; Carrano, 1999). This result highlights the reduced involvement of the radius in weight support, compared to the other bones that either display stable or increasing robustness during growth, following a more graviportal pattern of robustness in adults (Carrier, 1996; Christiansen, 2002; Kilbourne & Makovicky, 2012).

The femur is the only bone to grow more robust with age; however, qualitative comparisons indicate that this is due to an increase in diaphyseal circumference only. The femoral epiphyses appear to show no clear variation in robustness, whereas the middle of the diaphysis grows much thicker (Bader *et al.*, 2023). This robustness variation is associated with a more homogeneous distribution of the cortical thickness along the diaphysis, indicating a release of constraints on the microanatomy during growth.

## Comparison with other heavy animals

Elephants, as the heaviest living terrestrial animals, exhibit morphological and microanatomical adaptations in the long bones of their limbs to support their massive weight. Those adaptations are however not unique, but shared with several other large-bodied taxa. In elephants, the medullary area is filled with trabecular bone, providing increased load transmission along the bones. This characteristic is shared with other large-bodied taxa such as rhinos, hippo, and sauropods, in which it is assumed to reinforce load-bearing capacity (Lefebvre et al., 2023; Etienne, 2023) and probably also as a means to counterbalance buoyancy and walk on the bottom of bodies of water in the case of hippos (Houssaye *et al.*, 2021). However, whereas elephants exhibit a relatively thick cortex as compared to most terrestrial taxa (Houssaye *et al.*, 2018), it is much thinner than in rhinos, in which the long bones show a very thick cortex, with a marked increase in thickness close to the growth center. Our results are consistent with findings for the elephant-sized sauropod *Nigersaurus* by Lefebvre et al. (2023), who proposed that this difference in cortical thickening in association to large body mass was linked to different limb postures. The much thinner cortex of the long bones in elephants and sauropods is correlated with the more upright organization of their limbs. Our observations are also in accordance with elephants being, unlike rhinos, unable to trot and to gallop (Ren *et al.*, 2008), which implies less intense bending moments imposed on their bones, and hence a thick cortex such as observed in rhinos is not as important. In accordance, while cortical thickness varies between forelimb and hindlimb bones in rhinos (Etienne et al., 2021, Etienne, 2023), it is much more homogeneous between the limbs in elephants, reflecting the more similar involvement of the limbs in both propulsion and braking in the latter (Ren et al., 2010). The same pattern was observed in sauropods (Lefebvre *et al.*, 2023). Elephants thus exhibit a microanatomical structure in their long bones similar to sauropods, which probably reflects the similar biomechanical constraints acting on the limbs in columnar-limbed taxa.

The greater cortical thickness around the growth center observed in the heavier *L. africana* relative to *E. maximus* suggests an increased reliance on bone microanatomy to handle axial compression, possibly compensating for the overall slimmer morphology of the long bones in *L. africana* (Bader *et al.*, 2023). Conversely, *E. maximus* might rely more on the more robust morphology of their limb bones, with relatively larger diaphyses and wider epiphyses (Bader *et al.*, 2023), to handle the mechanical load associated with body weight support, thus releasing the need for a thick cortex and allowing for a larger medullary area.  In elephants, the weight support thus appears to be managed as much by posture, as by bone external morphology and microanatomy.

# CONCLUSION

The six limb long bones of extant elephants share global microanatomical patterns adapted to heavy weight-bearing. Both species exhibit bones with a relatively thick cortex and almost entirely filled with trabecular bone, as expected in graviportal animals for improved weight- bearing. Both species display highly anisotropic trabeculae, mainly parasagittally oriented in the diaphysis and metaphyses, and thus along the main direction of load transfer in animals whose limbs are nearly orthogonal to the ground during standing and walking. Beyond this general pattern, the six long bones do not contribute equally to movement and support, which is reflected in their microanatomy: body mass appears to be supported slightly more medially in the forelimb's stylopod, and more laterally in the hindlimb. We also observed changes in microanatomy during growth: calves exhibit very low anisotropy that quickly increases to support the growing weight, and associated biomechanical constraints. Conversely, the hourglass-shaped distribution of the compact cortex already occurs in juvenile specimens, even before birth, indicating the adaptive inhibition of cortical bone resorption from the earliest life stages. Adult specimens of *E. maximus* and *L. africana* differ slightly: the hourglass-shaped distribution of compact cortex is more pronounced in *L. africana*, possibly due to its heavier body associated with the relatively slimmer long bone morphology. Finally, although elephants share microanatomical adaptations with other heavy taxa such as rhinos and hippos, their relatively limited increase in cortical bone compactness and the rather homogeneous organization between the forelimb and hindlimb bones are more akin to the conditions observed in sauropods, in accordance with their shared columnarity and the specific orientation and distribution of loads along the bones, associated with an assumed release of constraints allowing more limited reinforcement of the bones' inner structure.

# ACKNOWLEDGEMENTS

We warmly thank J. Lesur, C. Bens, A. Verguin, G. Billet (Muséum national d'Histoire naturelle (MNHN), Paris, France) and Louise Tomsett and Richard Sabin of the Natural History Museum, London for granting access to the specimens. Further thanks to M. Bellato from the AST-RX platform in the MNHN (UMS 2700, Paris, France) for performing scans and reconstructions. We also thank M. Amari, P. Costes, N. Mohanty, I. Pelletan, M. Sowinski, K. Stilson and C. Tetaert (MNHN, Paris, France) for their assistance in transporting heavy material, as well as Sandra Shefelbine and Michael Doube (previously at Imperial College, London) for further support.

Further thanks to two anonymous reviewers that provided helpful comments and improved the manuscript, as well as Jeff Streicher for editorial work.

## CREDIT STATEMENT

**Conceptualization** CB, AH; **Data curation** CB, AH, RG, VH, JH; **Formal Analysis** CB; **Funding Acquisition** AH; **Investigation** CB, AH, JH, VH; **Methodology** CB, AH; **Project Administration** CB, AH; **Supervision** AH; **Validation** CB, AH, RG, JH, VH; **Visualization** CB; **Writing – Original Draft Preparation** CB**; Writing – Review & Editing** CB, AH, JH, VH.

## FUNDING

This work was funded by the European Research Council as part of the GRAVIBONE project (ERC-2016-STG-715300). JRH was funded by grants from Biological Sciences Research Council (grants BB/C516844/1 and BB/H002782/1). The funders had no role in study design, data collection and analysis, decision to publish, or preparation of the manuscript.

## CONFLICT OF INTEREST

The authors declare that they have no conflicts of interest.

## DATA AVAILABILITY

Raw CT-scan data of the MNHN specimens are archived at the Muséum national d'Histoire naturelle, Paris, France and registered on the 3Dtheque portal: https://3dtheque.mnhn.fr/. Raw CT-scan data of the NHMUK and RVC specimens are registered on figshare: doi:10.6084/m9.figshare.26779570.

# SUPPLEMENTARY DATA

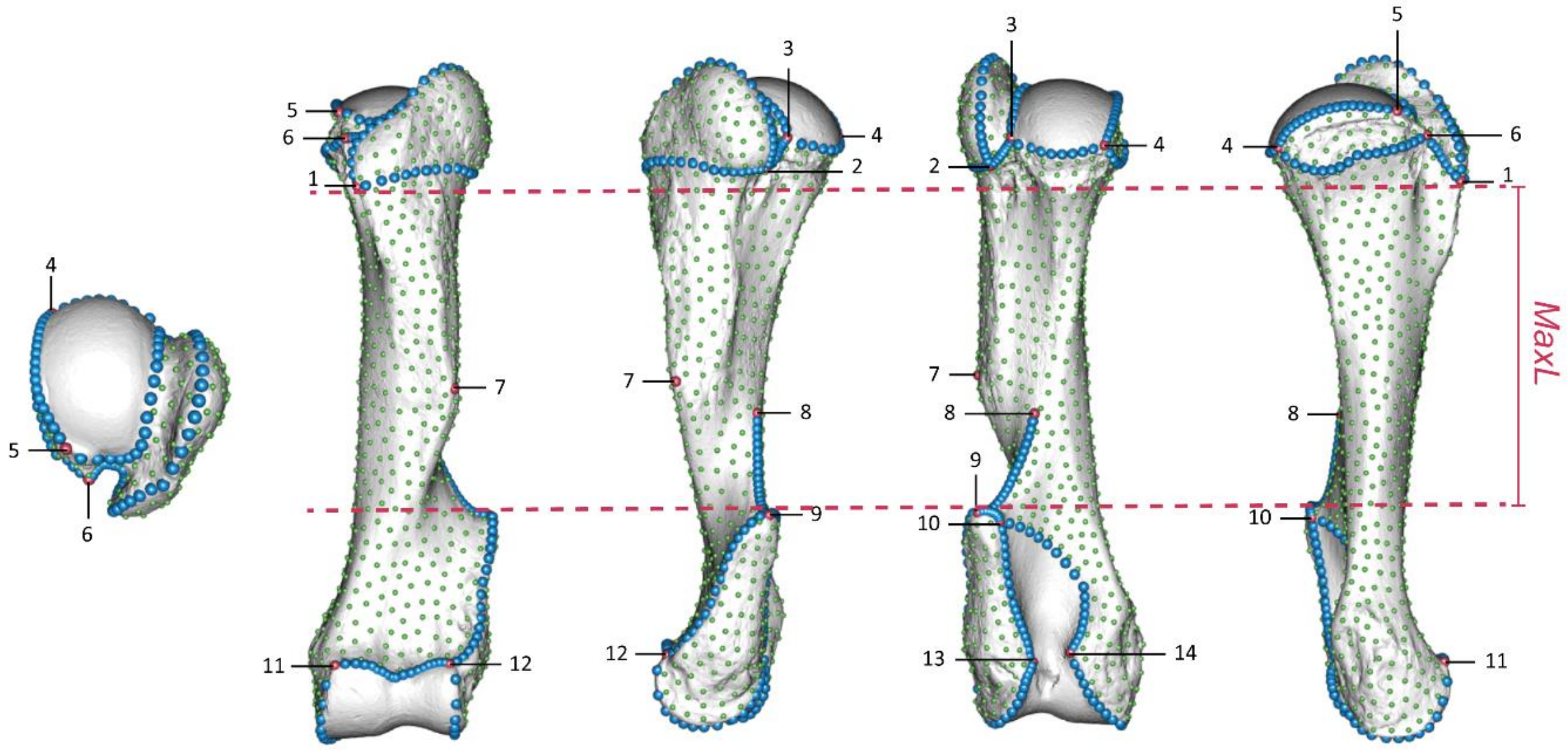

**Figure S1.** Location of the cutting planes for diaphyseal measures on the humerus. Landmarks following the methodology of Bader *et al*. (2024). *MaxL*, maximal diaphyseal length.

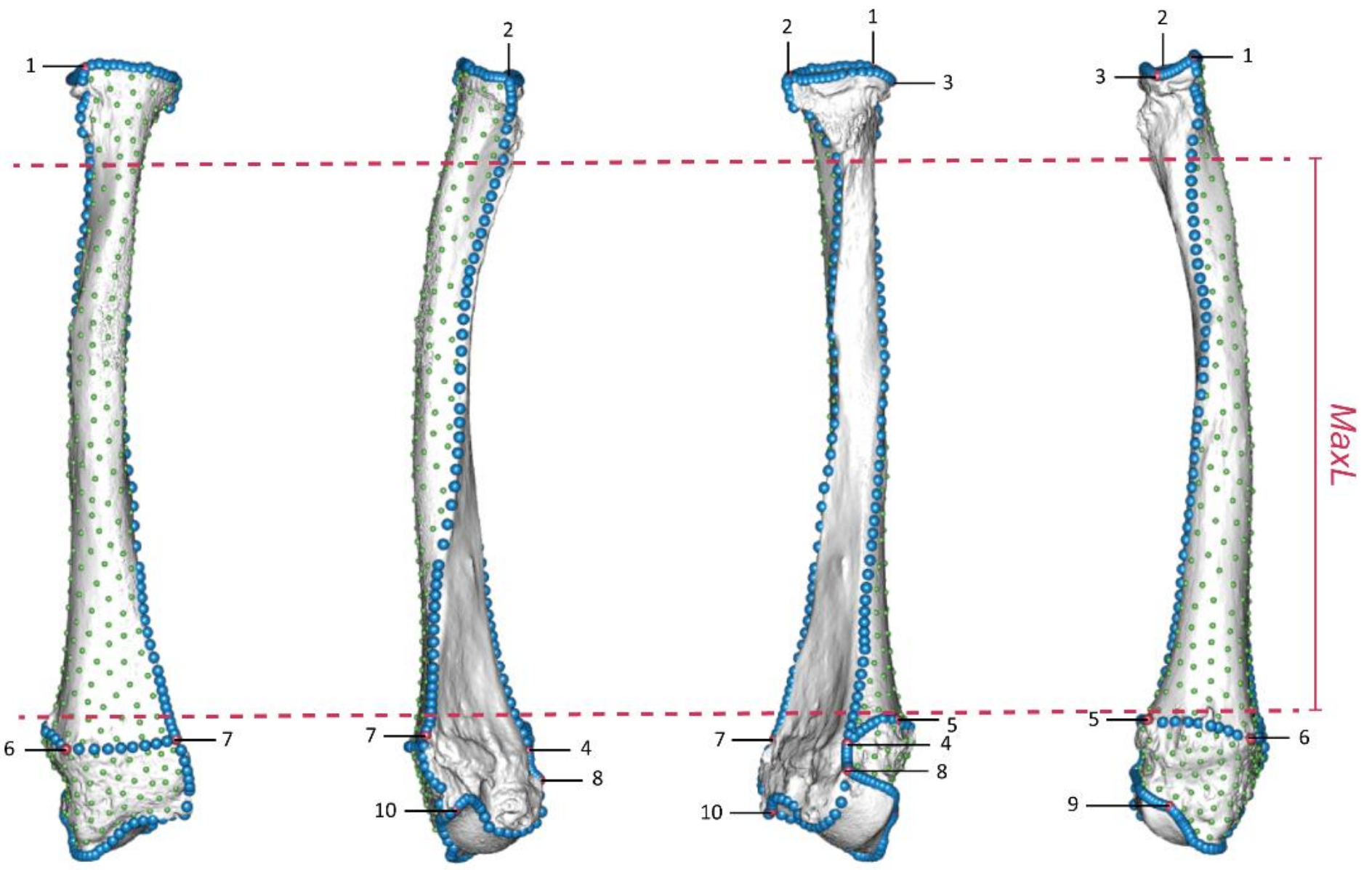

**Figure S2.** Location of the cutting planes for diaphyseal measures on the radius. Landmarks following the methodology of Bader *et al*. (2024). *MaxL*, maximal diaphyseal length.

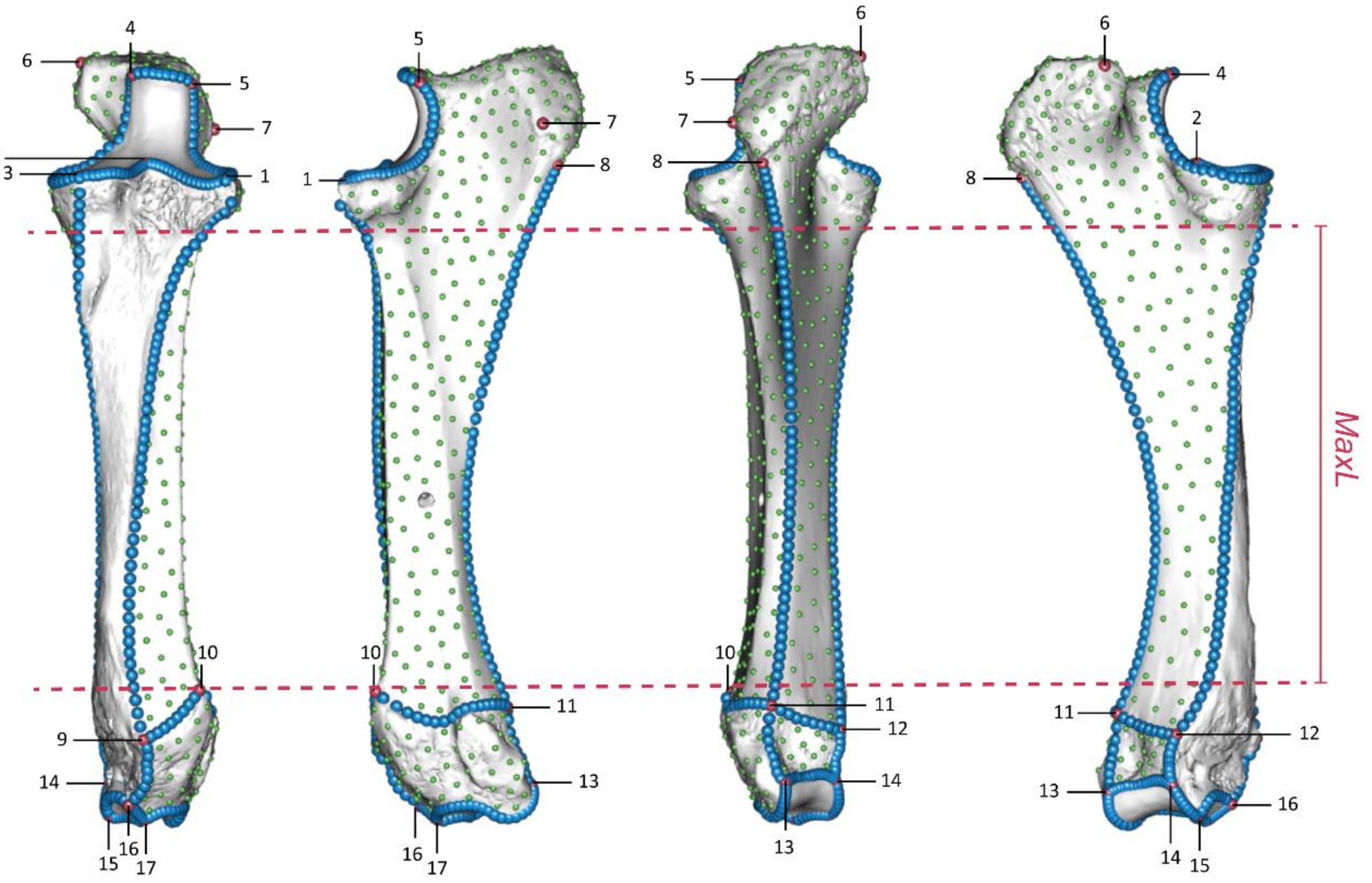

**Figure S3.** Location of the cutting planes for diaphyseal measures on the ulna. Landmarks following the methodology of Bader *et al*. (2024). *MaxL*, maximal diaphyseal length.

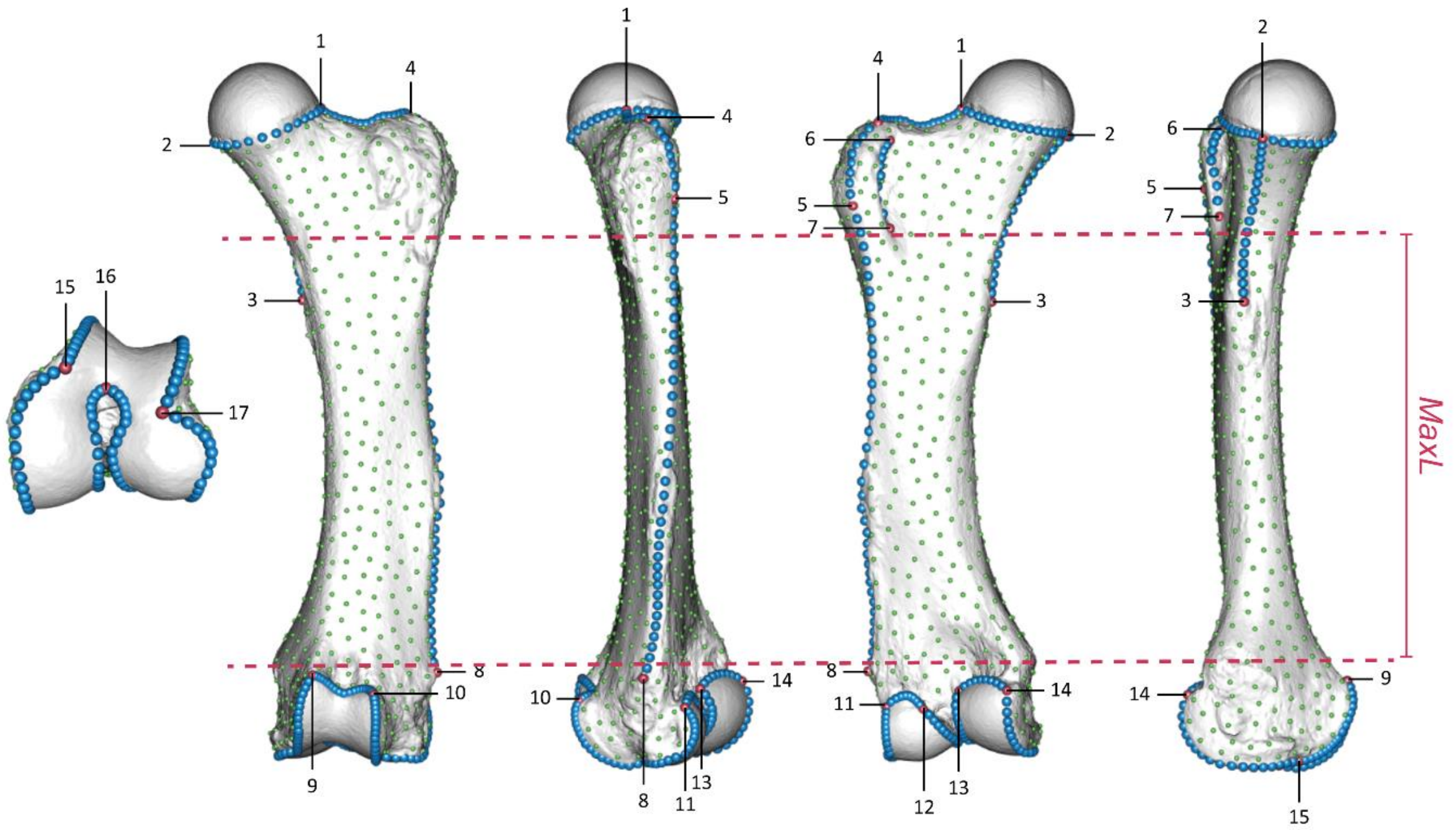

**Figure S4.** Location of the cutting planes for diaphyseal measures on the femur. Landmarks following the methodology of Bader *et al*. (2024). *MaxL*, maximal diaphyseal length.

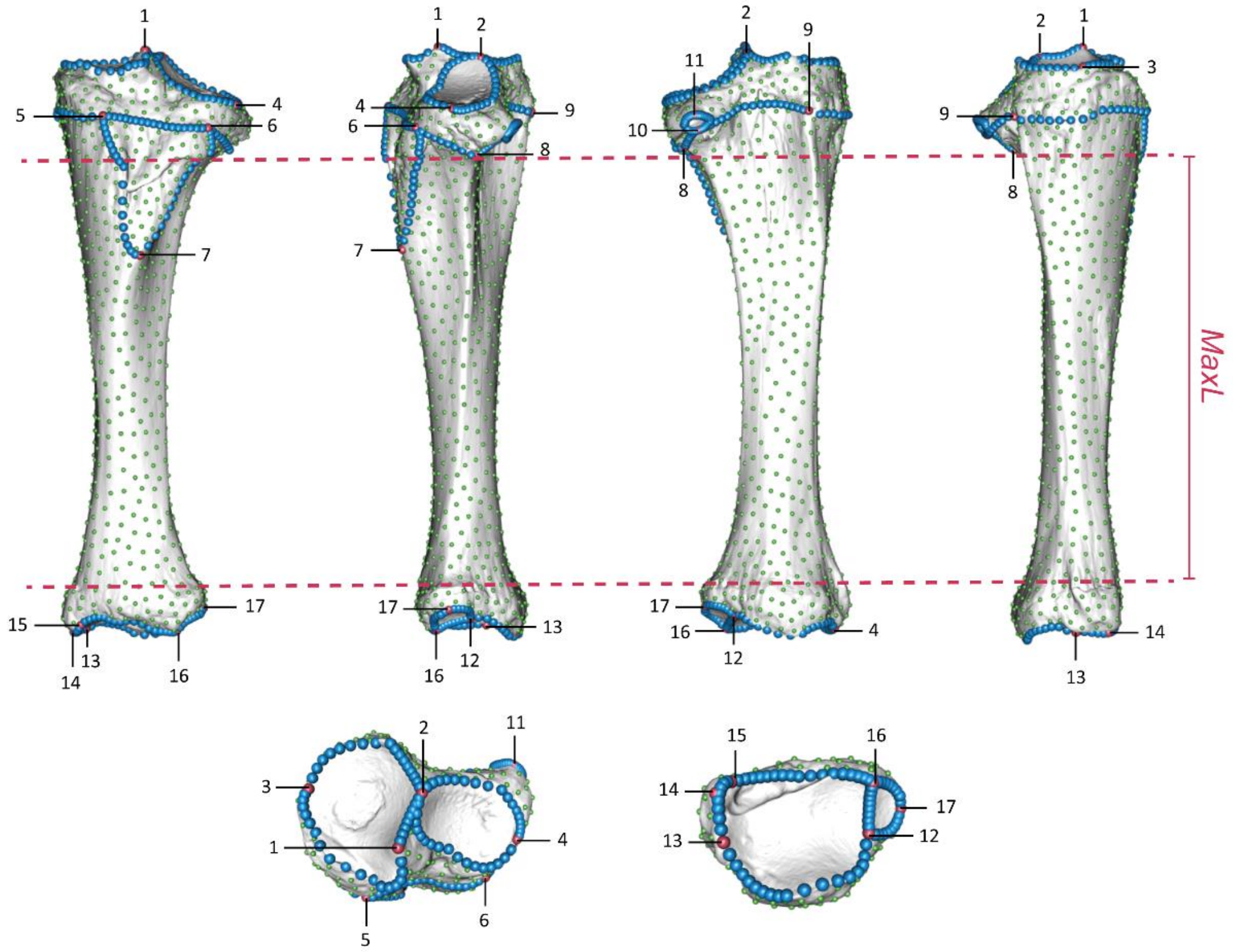

**Figure S5.** Location of the cutting planes for diaphyseal measures on the tibia. Landmarks following the methodology of Bader *et al*. (2024). *MaxL*, maximal diaphyseal length.

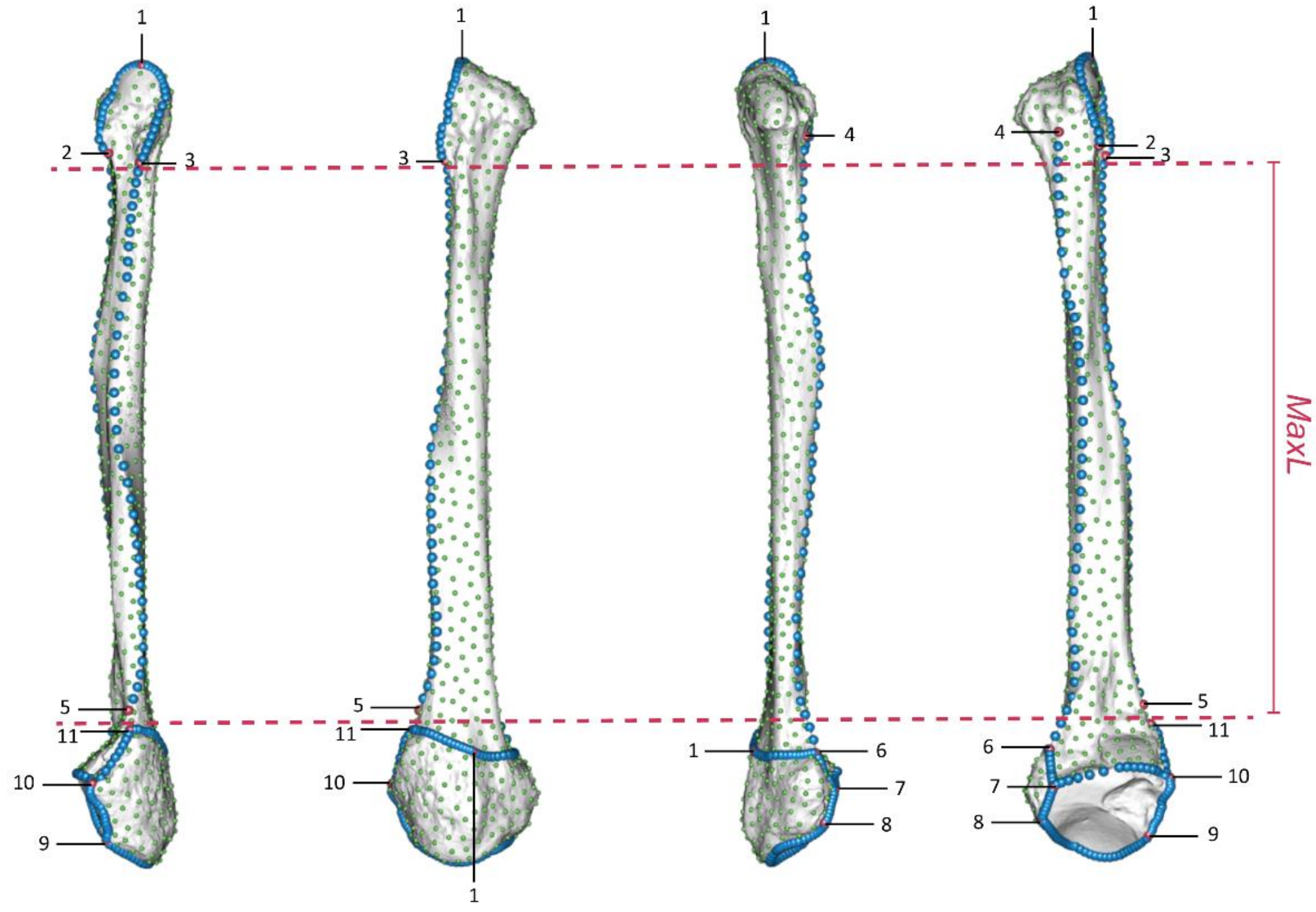

**Figure S6.** Location of the cutting planes for diaphyseal measures on the fibula. Landmarks following the methodology of Bader *et al*. (2024). *MaxL*, maximal diaphyseal length.

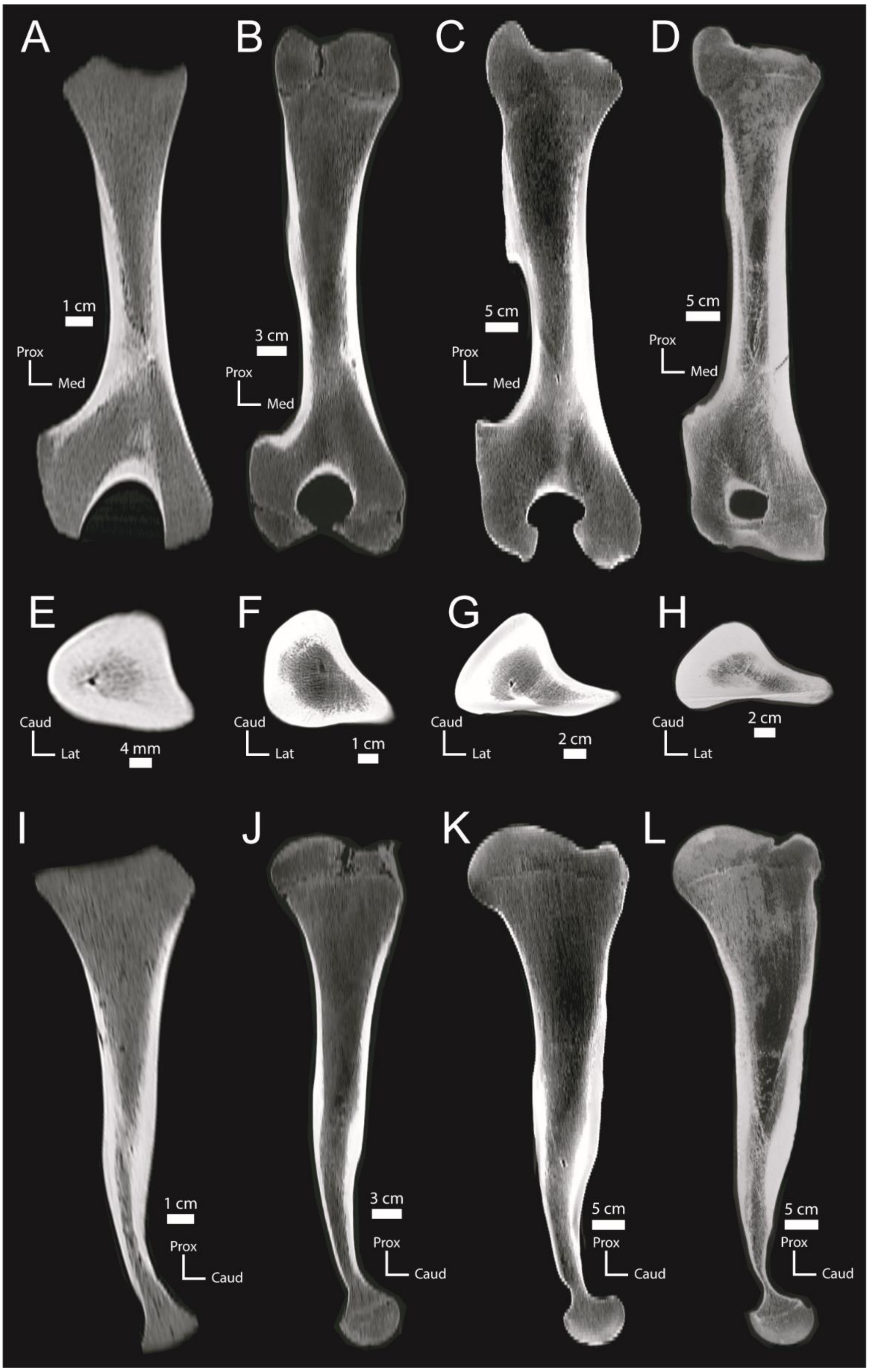

**Figure S7.** Virtual slices of the humerus *Elephas maximus* specimens. (A), (E), (I) calf (BMNH-1915.5.1.1), (B), (F), (J) subadult (BMNH-1984.516), (C), (G), (K) adult (BMNH-1907.3.18.1), (D), (H), (L) adult (BMNH-GNR) in (A), (B), (C), (D) coronal, (E), (F), (G), (H) transversal and (I), (J), (K), (L) sagittal view. Caud, caudal, Lat, lateral, Med, medial, Prox, proximal.

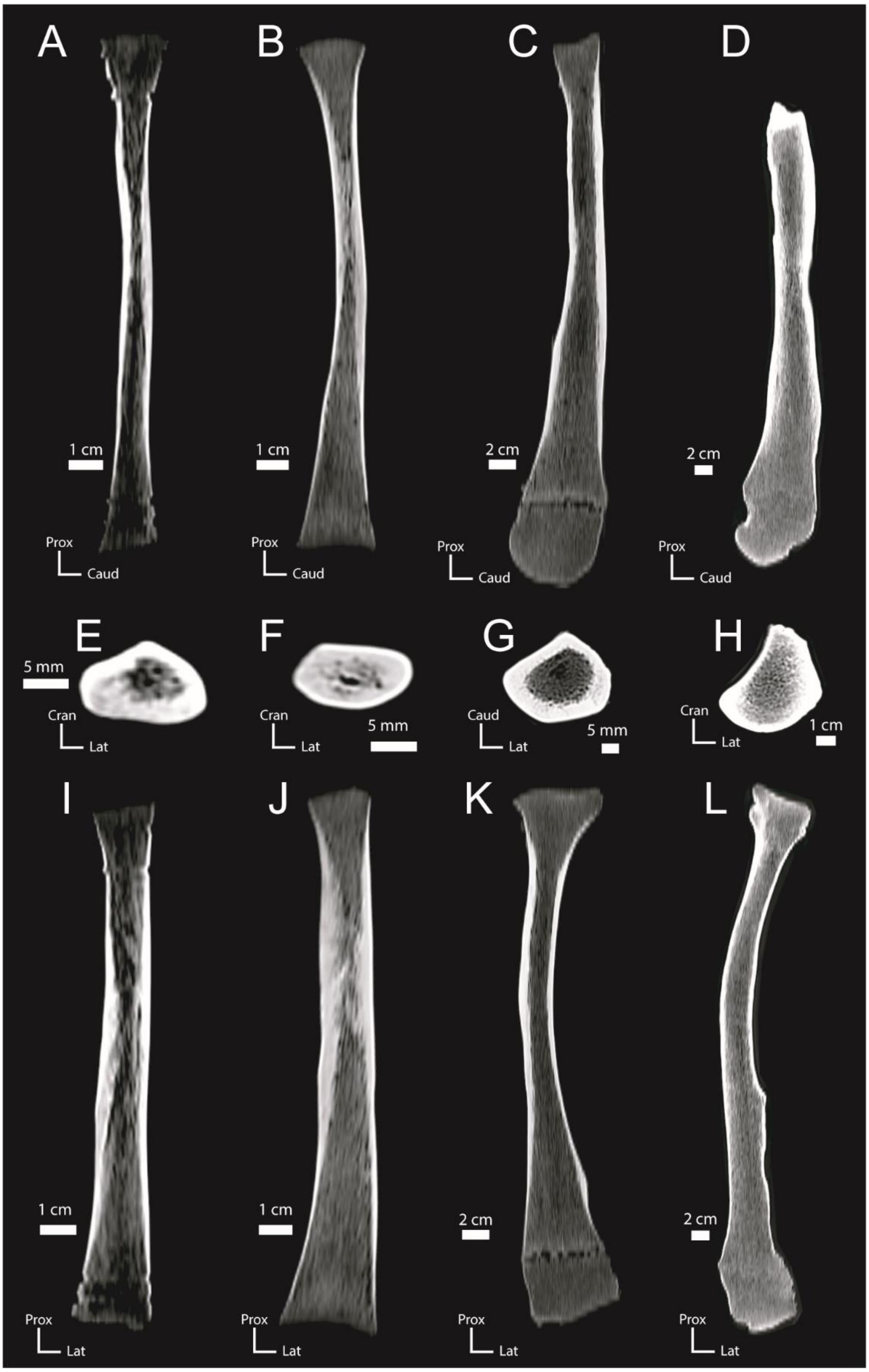

**Figure S8.** Virtual slices of the radius of (A), (E), (I) a fetus of *Loxodonta africana* (BMNH-1984.514), and *Elephas maximus* specimens: (B), (F), (J) calf (BMNH-1915.5.1.1), (C), (G), (K) subadult (BMNH-1984.516), (D), (H), (L) adult (BMNH-1907.3.18.1) in (A), (B), (C), (D) coronal, (E), (F), (G), (H) transversal and (I), (J), (K), (L) sagittal view. Caud, caudal, Lat, lateral, Prox, proximal.

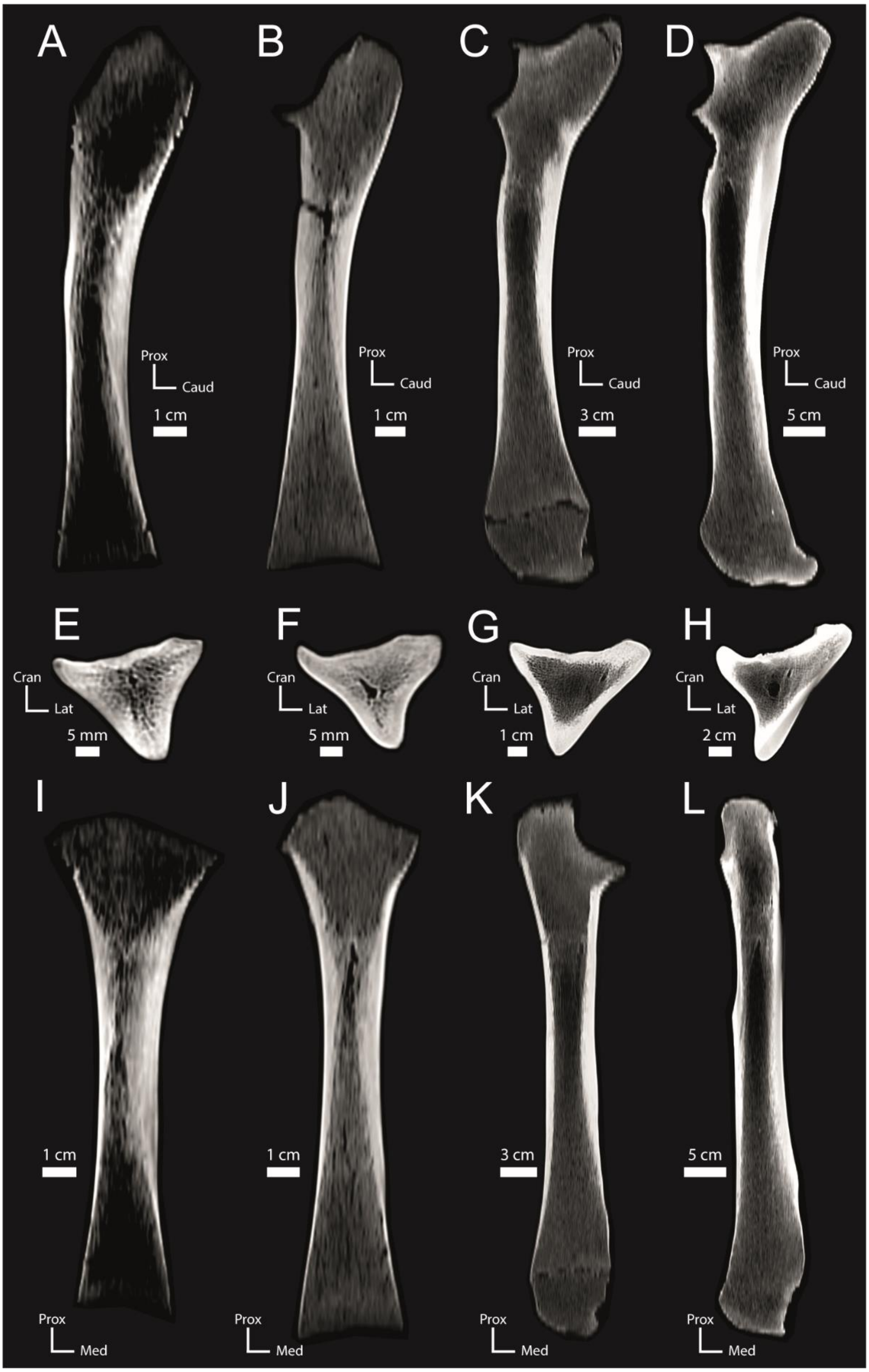

**Figure S9.** Virtual slices of the ulna of (A), (E), (I) a fetus of *Loxodonta africana* (BMNH-1984.514), and *Elephas maximus* specimens: (B), (F), (J) calf (BMNH-1915.5.1.1), (C), (G), (K) subadult (BMNH-1984.516), (D), (H), (L) adult (BMNH-1907.3.18.1) in (A), (B), (C), (D) coronal, (E), (F), (G), (H) transversal and (I), (J), (K), (L) sagittal view. Caud, caudal, Lat, lateral, Med, medial, Prox, proximal.

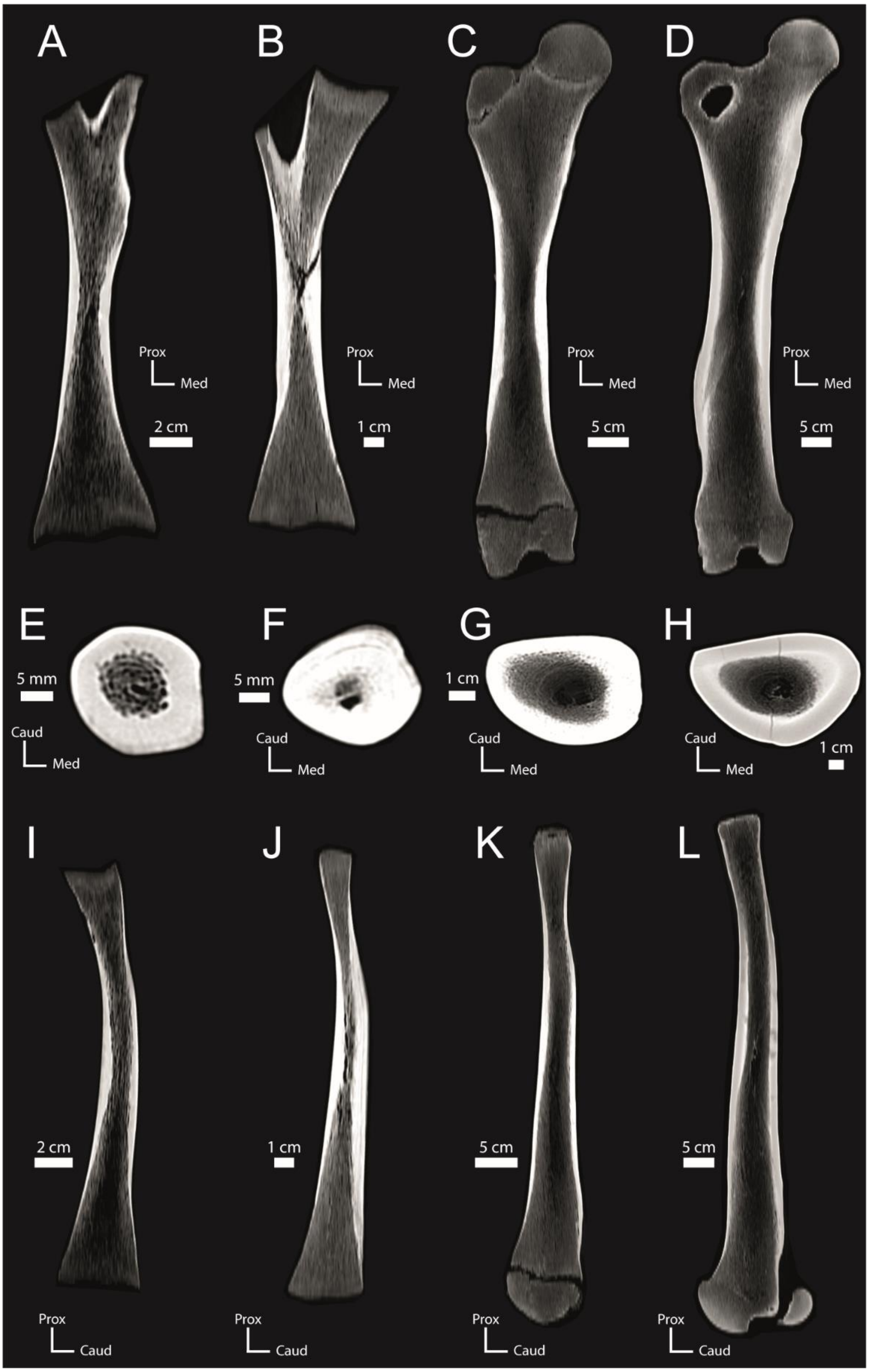

**Figure S10.** Virtual slices of the femur of (A), (E), (I) a fetus of *Loxodonta africana* (BMNH-1984.514), and *Elephas maximus* specimens: (B), (F), (J) calf (BMNH-1915.5.1.1), (C), (G), (K) subadult (BMNH-1984.516), (D), (H), (L) adult (BMNH-1907.3.18.1) in (A), (B), (C), (D) coronal, (E), (F), (G), (H) transversal and (I), (J), (K), (L) sagittal view. Caud, caudal, Prox, proximal.

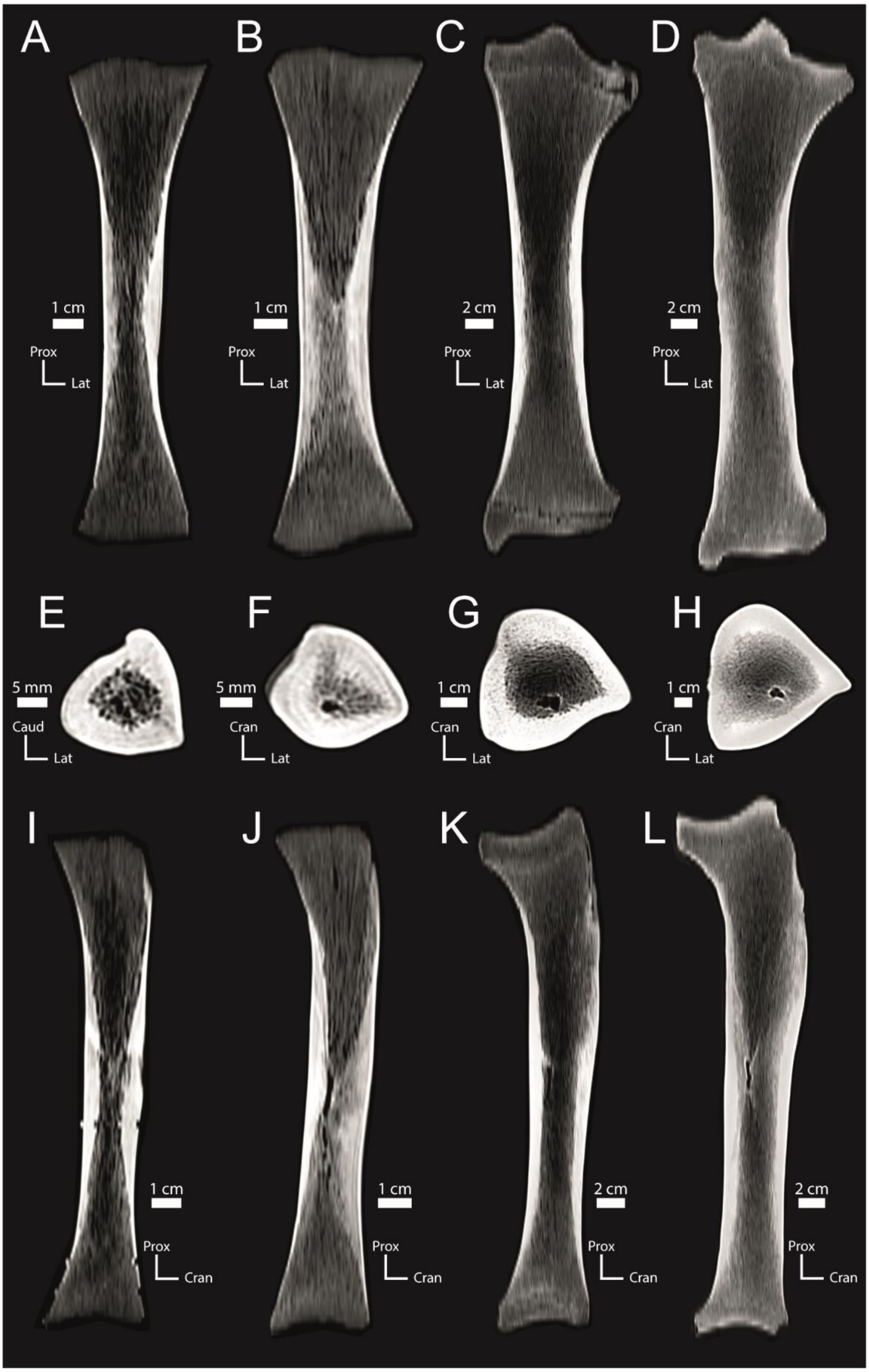

**Figure S11.** Virtual slices of the tibia of (A), (E), (I) a fetus of *Loxodonta africana* (BMNH-1984.514), and *Elephas maximus* specimens: (B), (F), (J) calf (BMNH-1915.5.1.1), (C), (G), (K) subadult (BMNH-1984.516), (D), (H), (L) adult (BMNH-1907.3.18.1) in (A), (B), (C), (D) coronal, (E), (F), (G), (H) transversal and (I), (J), (K), (L) sagittal view. Cran, cranial, Lat, lateral, Prox, proximal.

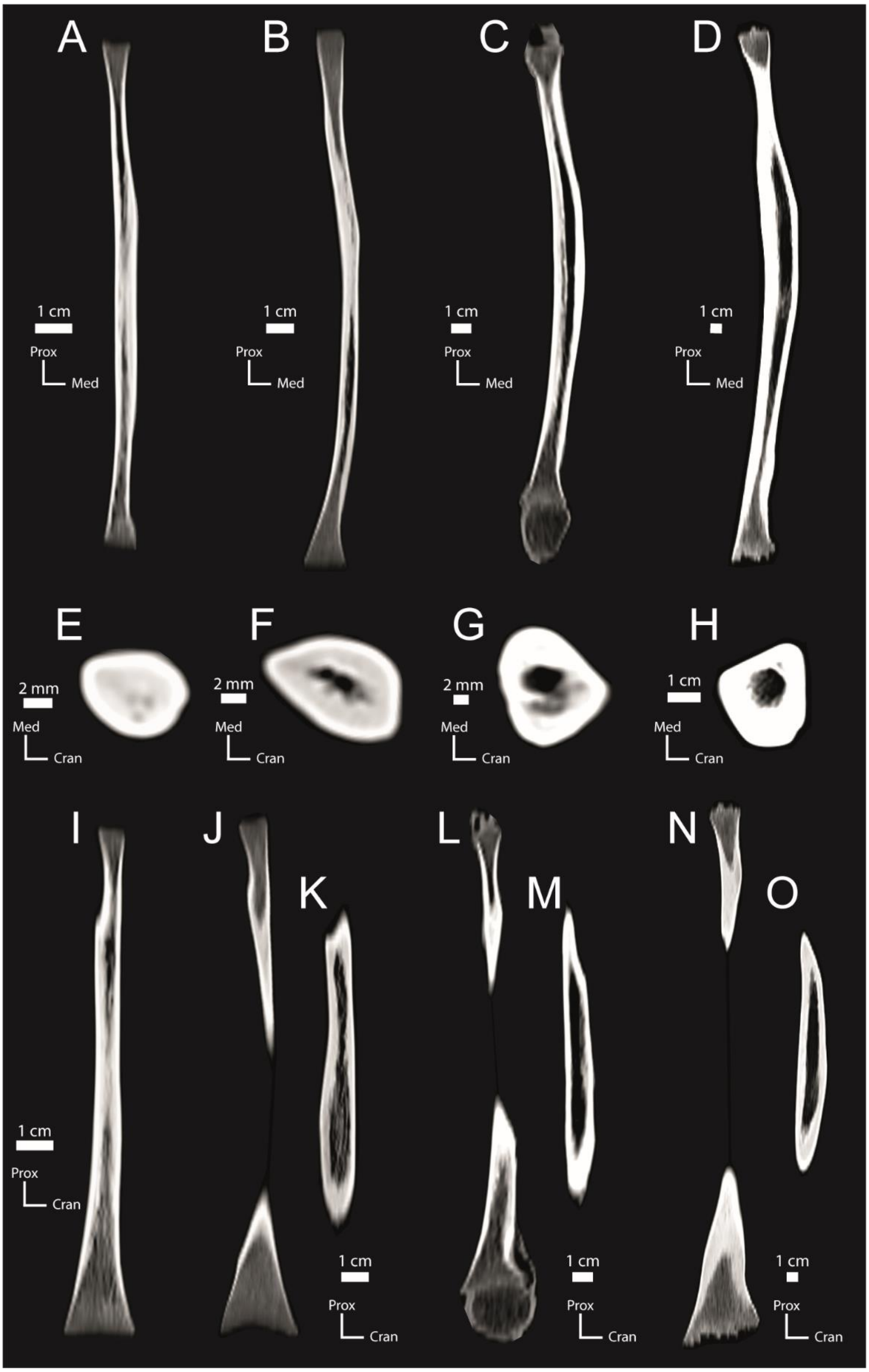

**Figure S12.** Virtual slices of the fibula of (A), (E), (I) an *Elephas maximus* calf (BMNH-1915.5.1.1), and *Loxodonta africana* specimens: (B), (F), (J) neonate (NHMUK-1962.7.6.8), (C), (G), (K) calf (BMNH-1962.7.6.9), (D), (H), (L) subadult (BMNH-1961.8.9.82) in (A), (B), (C), (D) coronal, (E), (F), (G), (H) transversal and (I), (J), (K), (L) sagittal view. Cran, cranial, Med, medial, Prox, proximal.

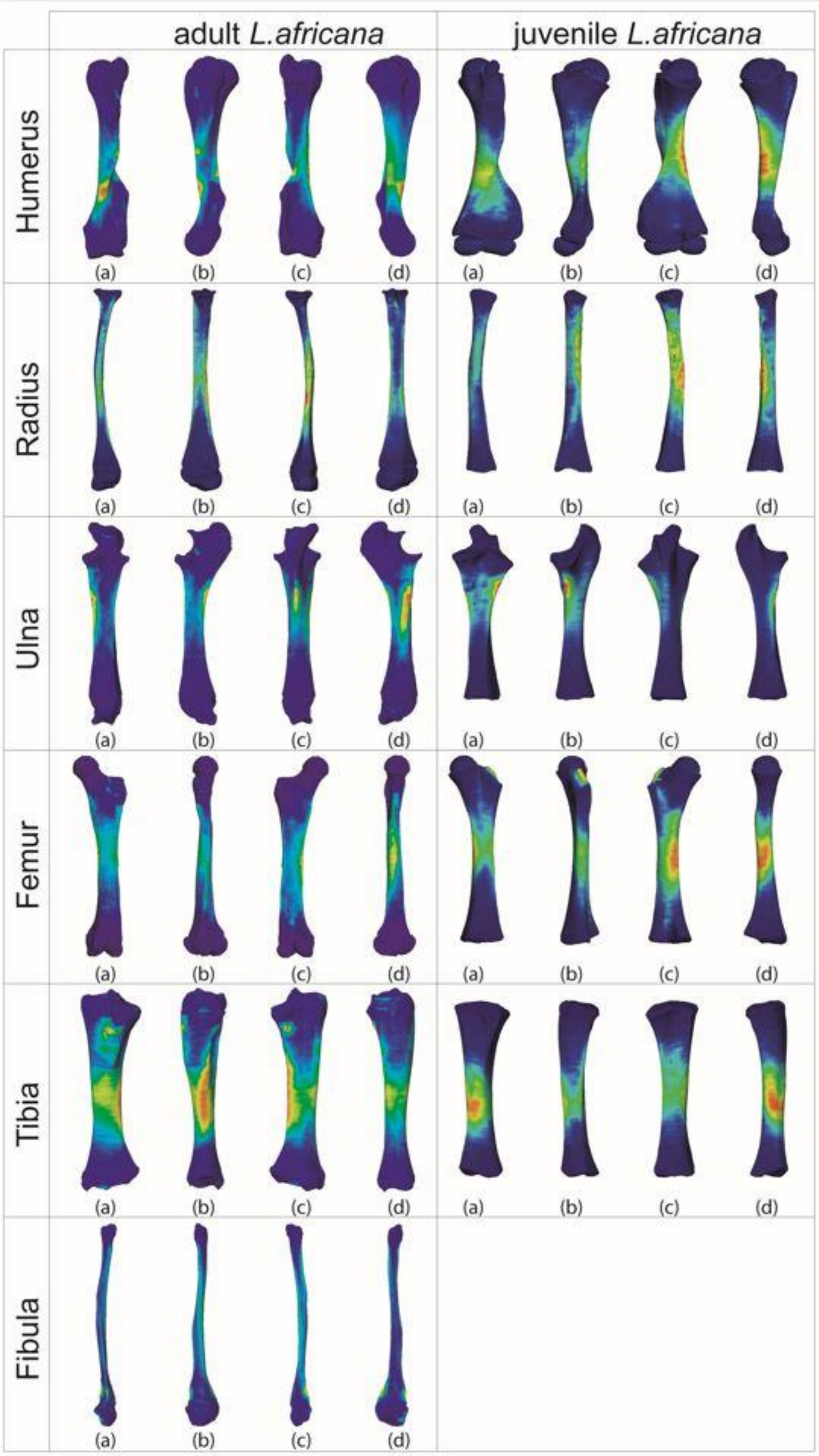

**Figure S13.** Visualizations of the cortical thickness mapping of the limb long bones of adult and juvenile specimens of *Loxodonta africana* in (a) cranial, (b) lateral, (c) caudal and (d) medial view. Cortical thickness is represented by a gradient ranging from cold (low cortical thickness) to warm (high cortical thickness) colors on the 3D mapping; values are relative to the minimum and maximum cortical thickness for each bone.

**Material S1**: 3D-mapping of the thickness of the outer layer of compact bone provides graphical outputs allowing for qualitative comparisons of the compact bone distribution. In order to obtain these 3D thickness maps, compact and trabecular bone need to be separated. Following Bader et al. (2022), we manually isolated an outer surface (corresponding to the outer surface of the bone) and an inner surface (corresponding to the inner limit of the compact cortex) for each bone, using Avizo 9.1. We then generated 3D bone cartographies using the 'SurfaceDistance' module in Avizo, i.e., calculating the thickness of cortical bone by measuring the distance between the outer and the inner surfaces of the cortex, and generating 3D cortical thickness maps of the entire bones using relative (i.e., normalized to the bone minimum and maximum) values. The resulting 3D mappings were used to visually assess the patterns of cortical thickness distribution along the bones.